\documentclass[twocolumn,showpacs,preprintnumbers,amsmath,amssymb,showkeys]{revtex4}

\usepackage[]{graphicx}
\usepackage[]{color}

\renewcommand{\d}{{\rm d}}
\newcommand{\fc}{{f_{\rm c}}}

\newcommand{\fNyq}{{f_{\rm Nyq}}}
\newcommand{\fsample}{{f_{\rm sample}}}

\newcommand{\Pbarexp}{\bar{P}^{\rm (ex)}}
\newcommand{\Pbarexpq}{\bar{P}^{{\rm (ex)}q}}
\newcommand{\Pbarexpto}[1]{\bar{P}^{{\rm (ex)}#1}}
\newcommand{\Pexp}{P^{\rm (ex)}}

\newcommand{\lsq}{^{\rm (lsq)}}
\newcommand{\kT}{k_{\rm B} T}
\newcommand{\Tmsr}{t_{\mbox{\scriptsize msr}}}
\renewcommand{\Re}{{\rm Re}}
\renewcommand{\Im}{{\rm Im}}

\newcommand{\Ftherm}{F_{\rm therm}}
\newcommand{\Dlsq}{D^{\rm (lsq)}}

\newcommand{\atrick}{\hat{a}}
\newcommand{\btrick}{\hat{b}}
\newcommand{\ctrick}{\hat{c}}
\newcommand{\Atrick}{\hat{A}}
\newcommand{\Btrick}{\hat{B}}

\newcommand{\beq}{\begin{equation}}
\newcommand{\eeq}{\end{equation}}
\newcommand{\bea}{\begin{eqnarray}}
\newcommand{\eea}{\end{eqnarray}}
\newcommand{\e}{\enspace}
\newcommand{\la}{\langle}
\newcommand{\ra}{\rangle}
\newcommand{\LA}{\left<}
\newcommand{\RA}{\right>}

\begin{document}
\title{Power spectrum analysis with least-squares fitting:
Amplitude bias and its elimination, with application to optical tweezers and atomic force microscope cantilevers}

\date{\today}

\author{Simon F. N\o{}rrelykke\\
{\small Max Planck Institute for the Physics of Complex Systems, 01187 Dresden , Germany.}\\
 {\small Department of Molecular Biology, Princeton University, Princeton, New Jersey 08544, USA.}\\ 
 and \\
 Henrik Flyvbjerg\\
 {\small Department of Micro- and Nanotechnology, Technical University of Denmark,
2800 Kongens Lyngby, Denmark.}}

\begin{abstract}
Optical tweezers and AFM cantilevers are often calibrated
by fitting their experimental power-spectra of Brownian motion.
We demonstrate here that if this is done with typical weighted least-squares methods
the result is a bias of relative size between $-2/n$ and $+1/n$ 
on the value of the fitted diffusion coefficient.
Here $n$ is the number of power-spectra averaged over,
so typical calibrations contain 10--20\% bias.
Both the sign and the size of the bias depends on the weighting scheme applied.
Hence, so do length-scale calibrations based on the diffusion coefficient.
   
The fitted value for the characteristic frequency is not affected by this bias.
For the AFM then, force measurements are not affected provided an independent length-scale calibration is available. 
For optical-tweezers there is no such luck, since the spring constant is found as the ratio of the characteristic frequency and the diffusion coefficient.

We give analytical results for the weight-dependent bias for the wide class of systems whose dynamics is described by 
a linear (integro-)differential equation with additive noise, white or colored.
Examples are optical tweezers with hydrodynamic self-interaction and aliasing, calibration of Ornstein-Uhlenbeck models in finance, models for cell-migration in biology, etc.
Because the bias takes the form of a simple multiplicative factor on the fitted amplitude (e.g.\ the diffusion coefficient) it is straightforward to remove, and the user will need minimal modifications to his or her favorite least-square fitting programs.

Results are demonstrated and illustrated using synthetic data,
so we can compare fits with known true values. 
We also fit some commonly occurring power spectra \emph{once-and-for-all} 
in the sense that we give their parameter values and associated error-bars
as explicit functions of experimental power-spectral values.
\end{abstract}


\maketitle

\section{Introduction}
Optical tweezers (OTs) are often calibrated by fitting 
the power spectrum of a trapped bead's Brownian motion with a Lorentzian~\cite{Neuman2004,Neuman2007,Neuman2008,Moffitt2008,Perkins2009}.
Similarly, atomic force microscopes (AFMs) are sometimes calibrated 
by fitting the power spectrum of the cantilever's Brownian motion 
with the power spectrum of a damped harmonic oscillator~\cite{Hutter1993,Walters1996,Sader1998}.
These fits are routinely done by least-squares (LSQ) minimization,
the premises of which are rarely satisfied in practice.
Here we do Maximum Likelihood Estimation (MLE) of parameters,
the premises of which \emph{are} satisfied,
and give the parameter values, with error bars, 
as explicit functions of the experimental power spectrum in an excellent approximation.
These closed-formula results should be useful for on-line calibration,
since that requires high-speed determination of parameters.
They also demonstrate that typical calibrations using least-squares fitting contain 10--20\% systematic errors, also known as \emph{bias}.

We find quite generally that both the sign and the size of the bias depends on the details of how a least-squares fit is implemented through the choice of how the data-points are weighted.  
Analytical expressions for the bias are given and examples of the behavior of the stochastic fit-errors are examined numerically.

These results apply beyond the examples given here, 
since the same bias occurs in all systems 
described by a linear (integro-)differential equation with additive noise.
Thus, using the correction factors given in Eqs~(\ref{eq:Pbias})~and~(\ref{eq:Pbiastheo}), it is possible to obtain correct unbiased estimates for the fit parameters without performing a computationally expensive MLE\@. Instead, one simply adjusts the results of an ordinary least-squares fit.

We also show that the bias found for least-squares fits appears under fairly general conditions, independent of the details of the function that is fitted, but depending only on the distribution of the error on the data. 
Analytical correction factors are given for the examples of Gaussian, exponential, and gamma distributed errors.
Finally, a general criterion is given for when one may expect a least-squares fit to be biased: 
When the experimental mean is correlated with the experimental variance.  Conversely, if they are uncorrelated the fit is unbiased.

The paper is organized as a main text followed by six appendices that contain proofs and other technical details.

\section{Dynamics}
The equation of motion for a massive particle moving in a harmonic potential under the influence of thermal forces is
\beq  \label{eq:NewtonMain}
	m\ddot{x}(t) + \gamma \dot{x}(t) + \kappa x(t) = \Ftherm(t) \enspace.
\eeq
Here $x(t)$ is the coordinate of the particle as function of time $t$, $m$ its inertial mass, $\gamma$ its friction coefficient, $\kappa$ is Hooke's constant, and $\Ftherm$  is the thermal force on the particle.
This force is random, and assumed to have white-noise statistical properties,
\bea  \label{eq:whitenoise}
	\langle \Ftherm(t) \rangle &=& 0 \\
	\langle \Ftherm(t) \Ftherm(t')\rangle &=& 2\kT \gamma \, \delta(t-t'),\,  \mbox{for all } t, t' \nonumber\e,
\eea
where $\delta$ is Dirac's delta function, $\kT $ the Boltzmann energy, and $\la \cdot \ra$ is the expectation value with respect to  the noise.
This theory's power spectrum of thermal motion is derived in \cite{Berg-Sorensen2004}.

\section{Power spectra}
Optical tweezers should be calibrated using the hydrodynamically correct
power spectrum given in \cite{Berg-Sorensen2004,Tolic-Norrelykke2006RSI}, possibly taking into account
a number of effects listed there and further detailed in \cite{Berg-Sorensen2003,Berg-Sorensen2006,Schaeffer2007}.
These effects include:
(i) The frequency dependence of the friction coefficient, due to hydrodynamic self-coupling;
(ii) dependence of the friction coefficient on distance to nearby surfaces,  due to hydrodynamic coupling;
(iii) extra $1/f$ power at low frequencies, caused by the laser pointing stability, hydrodynamic self-coupling, etc.; and
(iv) optical interference effects, caused by a standing wave between the trapped object and the nearby  microscope cover-slip surface, when determining the displacement sensitivity (Volt to nano-meter conversion factor).
Generally, the most crucial step, where the largest systematic errors are likely to appear, is the determination of the displacement sensitivity \cite{Tolic-Norrelykke2006RSI}.
However, as also described in \cite{Berg-Sorensen2004},
there are situations in which an acceptable approximation is achieved
by fitting a Lorentzian,
\beq  \label{eq:Lorentzian}
	P_f = \frac{D/(2\pi^2)}{\fc^2 + f^2} \enspace,
\eeq
to an average $\Pbarexp_f$ of, say, $n$ two-sided experimental power spectra
for the Brownian motion of a trapped microsphere (occasionally confusion arises over extra or missing factors of two in PSDs---this is down to the use of  one-sided, $f > 0$, versus two-sided, $f \gtrless 0$, PSDs; as long as the total power is contained in the chosen frequency range the two approaches are equally correct).
The notation used is essentially the same as in \cite{Berg-Sorensen2004}, where the aliased Lorentzian is also derived, i.e., $\fc = \kappa/(2\pi\gamma)$ and $D=\kT/\gamma$.

Similarly, AFM cantilevers are sometimes calibrated by fitting a frequency interval around the resonance peak \cite{Sader1998} using
\bea \label{eq:AFM}
	P_f &=& \frac{D/(2\pi^2)}{(\frac{2\pi m}{\gamma})^2 ( f_0^2 - f^2)^2 + f^2} \\
	&=& \frac{D/(2\pi^2)}{(\frac{Q}{f_0})^2 ( f_0^2 - f^2)^2 + f^2} \e,
\eea
to a similar average of experimental power spectra for a cantilever's Brownian motion.
Here, the characteristic frequency $f_0^2 = \kappa/(4\pi^2 m)$ and the quality factor $Q=\sqrt{m \kappa}/\gamma$.
Describing the AFM cantilever as a simple harmonic oscillator is rigorously correct for high quality factors $Q \gg 1$, e.g.\ in air where dissipative forces are small \cite{Sader1998}.
In water, this description also works, at least for short stiff cantilevers, but now the drag coefficient is frequency dependent \cite{Walters1996}.
It is not our aim here to review the expansive literature on AFM calibration---the only point we seek to make, is that if the amplitude of the power spectrum is involved, systematic fitting errors are typically present as shown below.

Both these theoretical power spectra follow from the Einstein-Ornstein-Uhlenbeck theory
for the Brownian motion of a damped harmonic oscillator in one dimension; see  Eq.~(\ref{eq:NewtonMain})  and  \cite{Berg-Sorensen2004},
where the parameters in Eqs.~(\ref{eq:Lorentzian}) and (\ref{eq:AFM}) are also defined.
In Appendix~\ref{app:aliasedAFM} we give the results for the aliased AFM power spectral density (PSD)\@.

Least-squares fits of power spectra are biased, as shown below in Section~\ref{sec:LSQ}, irrespective of whether one applies
the above simplified theory or a more complete one that takes into account aliasing, hydrodynamics, electronic filters etc.
This bias will be on the amplitude \emph{only} and will not affect shape parameters, i.e., $D$ will be biased, whereas $\fc$, $f_0$, and $Q$ will not.

\section{Statistical properties of experimental power spectrum}
\label{app:statprop}
Here, and throughout the rest of the paper, we differentiate between theoretical \emph{expectation} values, indicated by brackets, $\la\cdot\ra$, and experimental \emph{averages}, indicated by a bar, $\bar{\cdot}$.
Following  the notation of \cite{Berg-Sorensen2004}, we write the thermal force in Eq.~(\ref{eq:whitenoise}) in terms of a normalized white-noise process $\eta(t)$ 
\beq
	\Ftherm(t) = \sqrt{ 2\kT \gamma } \, \eta(t) 
\eeq
whose Fourier transform obeys
\begin{equation}
	\langle \tilde{\eta}_k \rangle =0 ~;~~~
	\langle \tilde{\eta}^*_k \tilde{\eta}_{\ell} \rangle = \Tmsr\delta_{k,\ell}
	\label{eq:discnoise}
\end{equation}
where $k,\ell=-N/2+1, \ldots, N/2$ are integers.
Moreover, since $\eta(t)$ is an white-noise process, hence temporally uncorrelated, $\Re\, \tilde{\eta}_k$ and $\Im\, \tilde{\eta}_k$ are two mutually independent random variables, uncorrelated for different $k>0$, with Gaussian distributions by virtue of the Central Limit Theorem, or, equivalently, by virtue of $\eta(t)$ being the first derivative of a Wiener process with respect to time.
Consequently, the sum of their squares $|\tilde{\eta}_k|^2$ is a non-negative random variable,  uncorrelated for different $k>0$, with {\em exponential\/} distribution \cite{footnote}.
Hence, so are the experimental values $\Pexp_f$ for the power spectrum for $f=k/\Tmsr>0$.

Thus the dynamical theory defined in Eqs.~(\ref{eq:NewtonMain}) and (\ref{eq:whitenoise})
predicts not just the \emph{expectation value} $P_f$ for the experimental spectrum $\Pexp_f$, as given in Eq.~(\ref{eq:AFM}).
It predicts also the \emph{distribution} from which the experimental power spectrum is ``drawn": The power-spectral value $\Pexp_f$ at each frequency $f$ is an independent random number, drawn from an exponential distribution with expectation value $P_f$,
\beq
	p(\Pexp_f;P_f) = \frac{1}{P_f} \exp(-\Pexp_f/P_f)  \enspace.
\eeq

Consequently,
\beq
	\langle \Pexp_f \rangle = P_f  \enspace,
\eeq
\beq  \label{eq:sigmaLorentziantwo}
	\sigma(\Pexp_f)= \langle (\Pexp_f-P_f)^2 \rangle^{1/2} = P_f  \enspace,
\eeq
and  the signal-to-noise ratio $\langle \Pexp_f \rangle/\sigma(\Pexp_f)$ equals one.
This is why we average over $n$ experimental spectra,
\bea
	\Pbarexp_f \equiv \frac{1}{n}\sum_{i=1}^n \Pexp_{f,i} \e,
\eea 
before plotting and fitting: To reduce noise.  

If the $n$ spectra are statistically independent---as is the case if they are computed from data taken in non-overlapping time intervals---then we have, unchanged, that
\beq
	\langle \Pbarexp_f \rangle = P_f  \enspace,
\eeq
but
\beq \label{eq:Gammasigma}
	\sigma(\Pbarexp_f)= \sigma(\Pexp_f)/\sqrt{n} = P_f/\sqrt{n}  \enspace,
\eeq
and $\Pbarexp_f$ is distributed according to a distribution that is the convolution of $n$ identical exponential distributions, viz.\ the gamma-distribution:
\beq \label{eq:gammadist}
	p_n(\Pbarexp_f;P_f) = \Pbarexpto{n-1}_f \frac{ \left( n/P_f \right)^n  }{\Gamma(n) }
	\exp\left(-n\Pbarexp_f/P_f\right)  \enspace,
\end{equation}
where $n$ is the shape parameter, $P_f/n$ the scale parameter, and $\Gamma(n) = (n-1)!$.
The mean, $P_f$, is the product of the shape and the scale parameters, the mode is $P_f (n-1)/n$, the variance, $P_f^2/n$, is the product of the mean and the scale, and the skewness is $2/\sqrt{n}$.
From the known distribution it is easy to show that the $q$th moment of $\Pbarexp$ is
\beq \label{eq:qmoment}
	\la \Pbarexpq \ra = \frac{ \Gamma(n+q) }{ n^q \Gamma(n) } \, P_f^q \e,
\eeq
a result that we shall need later.

\begin{figure}[ht]
 \includegraphics[width=8cm]{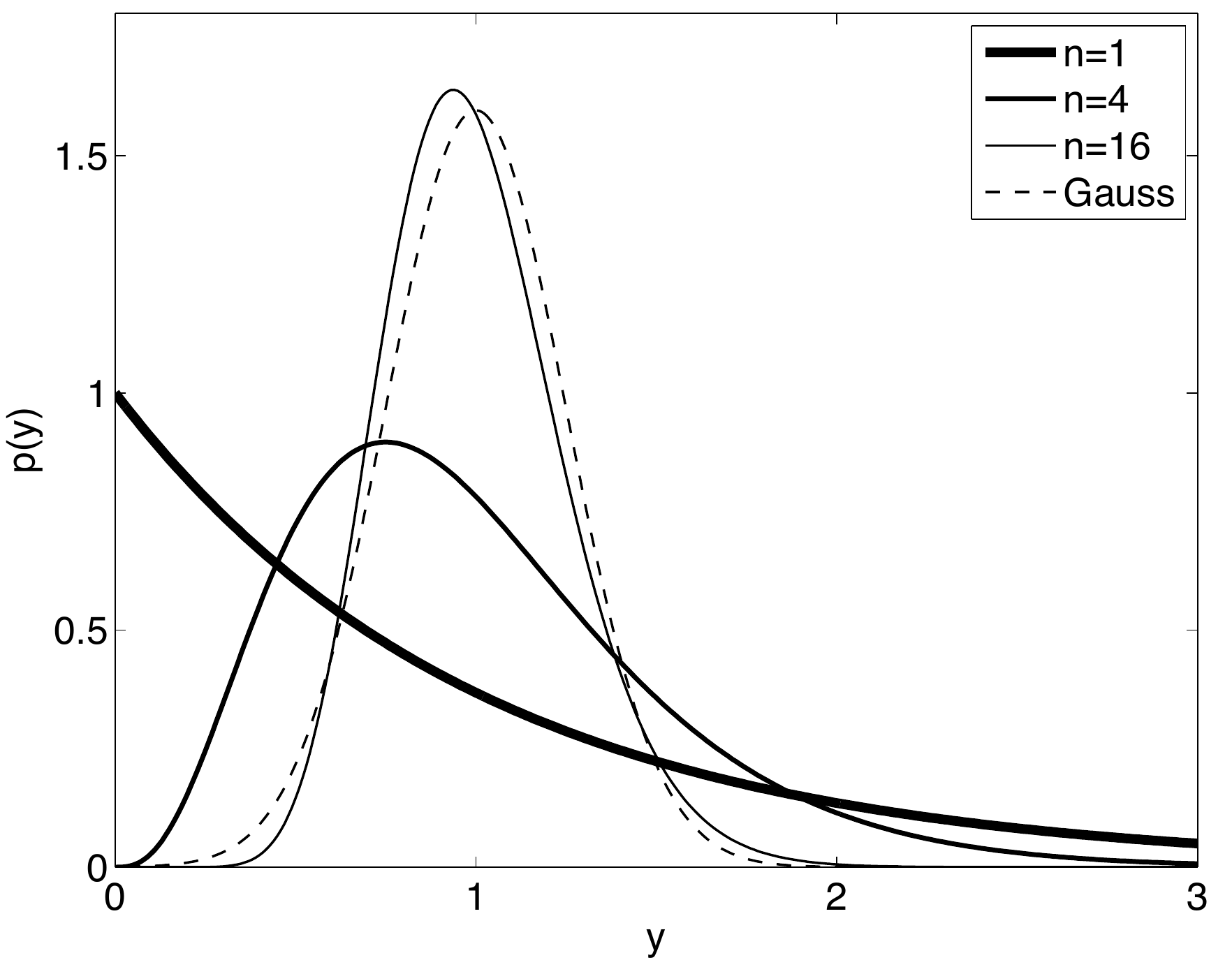}
\caption[]{\label{fig:pn}
The gamma distribution, Eq.~(\ref{eq:gammadist}),  as function of $\Pbarexp_f/P_f$ for various values of the shape parameter $n$:
Thick full line, $n=1$; full line, $n=2$; thin full line $n=16$.
Also shown is the Gaussian limit value, plotted here with a variance of $1/n=1/16$ (dashed line).}
\end{figure}

In the limit $n\rightarrow\infty$ this distribution approaches a Gaussian by courtesy of the Central Limit Theorem, but it is a slow approach, because the starting point, the exponential distribution, is highly skewed.
For moderate values of $n$, $p_n(\Pbarexp_f;P_f)$ is far from Gaussian, see Fig.~\ref{fig:pn}.
It is also quite skewed, but this is \emph{not} the source of the bias.
The bias is caused by the correlation between the experimental mean, $\Pbarexp_f$, and the experimental variance
\bea
	s_n^2(\Pbarexp_f) \equiv \frac{1}{n-1}\sum_{i=1}^n  (\Pexp_{f,i} - \Pbarexp_f)^2 \e,
\eea
as discussed in greater detail below.
The latter is easily shown to satisfy $\la s_n^2(\Pbarexp_f) \ra = \sigma^2(\Pbarexp_f)$.

\subsection{Ubiquity of the gamma-distribution}
\label{subsec:ubiq}
As a matter of fact, all systems described by a linear (integro-)differential equation with additive noise, be it white or colored, have power spectral values that are described by the statistics in Eq.~(\ref{eq:gammadist}).
To see that white and colored noise lead to the same statistics, all we need to realize is that the power spectral density of a colored noise is the product of the power spectral density of a white noise and some function $F(f)$ that describes the color of the noise (a filter function):
Quite generally, if the dynamics of $x$ is given by a polynomial $Q$ in $\d/\d t$ with constant coefficients 
\beq
	Q \left( \frac{\d}{\d t} \right) \, x(t) = \psi(t) \e,
\eeq
where $\psi$ is a colored noise, then the power spectrum 
\beq	
	 \Pexp_f (x) = \frac{ |\tilde{\psi}_f|^2 } { |Q(-i 2 \pi f)|^2 } = \frac{ |\tilde{\eta}_f|^2  F(f) }{ |Q(-i 2 \pi f)|^2}
\eeq
is seen to be the product of the power spectrum of a white noise $\eta$ and some function of $f$.
Thus, for a given frequency $f$, the statistics of the power spectral value is determined by the statistics of white noise.

\section{Least-square fitting}
\label{sec:LSQ}
As alluded to in the previous section, one effect of the exponential distribution of $\Pexp_{f,i}$ is that weighted least-square fitting to $\Pbarexp_f$ does not follow the text-book behavior and returns a result that is systematically wrong, or, biased. 
In the following we remind the reader of the rather narrow set of circumstances under which least-square fitting works.
We then relax the requirements and see what effect it has on the fit results.

Weighted least-squares (WLS) fitting minimizes
\beq 
	\label{eq:chi2}
	\chi^2 (\theta) = \sum_{i=1}^N \left[ y_i - f_i(\theta)\right]^2 w_i^2
\eeq 
with respect to fit-parameters $\theta$.
Here, $y_i = f_i + e_i$ are the experimental data (dependent variable), $e_i$ errors, $f_i$ the fit-function evaluated at $i$ (the independent variable), and $w_i$ are weights.
It is assumed that the experimental data are uncorrelated and that the weights are uncorrelated with the experimental data.  If, in addition, the weights are also independent of $i$, and thus constant, weighted least squares reduce to ordinary least squares (OLS). 
When the theoretical model, $f_i$, is a linear function of the fit parameters, $\theta$,  both OLS and WLS minimization are known to return parameter estimates that are BLUE (best linear unbiased estimate).
The former was first shown by Gauss in 1829 whereas the latter was shown by Aitken in 1935 \cite{Aitken1935}. Specifically, in WLS the errors do not have to be independent, just uncorrelated, and in OLS they do not have to come from the same distribution, they are only assumed to have the same variance (homoscedasticity). 

Things turn particularly simple when the experimental data are Gaussian distributed.
We can then choose $y_i$, e.g., as the average of several measurements, and the weights, $w_i$,  as the reciprocal of the experimental standard deviation.  
With this choice,  $y_i$ and $w_i$ are independent, because the sample mean and sample variance calculated from independent, identically Gaussian distributed random variables are statistically independent. 
In this Gaussian scenario it is possible to assign meaningful confidence intervals to the fit-parameters and calculate the goodness of the fit, also known as its \emph{support} or $p$-value, and this is why one is sometimes tempted to assume that data are Gaussian even when it is not quite the case.
Finally, with Gaussian data MLE and WLS are mathematically identical as it easily seen by inserting a Gaussian in $p_n$'s place in Eq.~(\ref{eq:pPexp}) and deriving the corresponding cost function, which is simply $\chi^2$.

\emph{We now ask what the effect is on the estimate of the fit-parameters if the data are not Gaussian,  the weights are not independent of the experimental value, and the model is a non-linear function of the fit parameters.}
All of these circumstances are realized when fitting  power spectra with the statistic given in the previous section.

\subsection{Some analytical results for bias in least squares fitting}
\label{subsec:analytic}
An estimator,  $\hat{\theta}$, is biased if its expectation value differs from the true value, $\la \hat{\theta} \ra \neq \theta^*$, of the quantity it is an estimator for.
Below, we show that some common least-square estimators for power spectra are biased.

At the minimum of the $\chi^2$ function given in Eq.~(\ref{eq:chi2}), the first derivative with respect to the fit parameters is zero, and the stationarity conditions thus reads
\beq \label{eq:genstat}
	\sum_{i=1}^N [ y_i - f_i] \, w_i^2 \, \partial_{\theta} f_i 
	= \sum_{i=1}^N [ y_i - f_i]^2 \, w_i \, \partial_{\theta} w_i \e.
\eeq
Notice, that in order to be completely general we have allowed the weights to depend on the fit parameters.
We now treat a few specific, but universal, scenarios one at a time.

\subsubsection{Experimental standard deviations as weights}
Proceeding as usual with classic WLS, we pick the weights to be inversely proportional to the experimental estimate for the standard deviation.  This estimate could be the known uncertainty from the experimental apparatus, but often it is estimated simply as 
\bea
	s_{n,i} =\sqrt{ \frac{1}{n-1}\sum_{j=1}^{n} (y_{i,j} -{\bar{y}_i})^2 }\e,
\eea 
where $n$ is the number of times the experiment is repeated with the independent variable set to its $i$th value. The experimental average (sample mean)
\bea 
	\bar{y}_i = \frac{1}{n}\sum_{j=1}^{n} y_{i,j}
\eea
is our estimate for the expectation value of $y_i$.
Doing this for each $i$, 
\beq 
	\label{eq:chi2sigma}
	\chi^2(\theta) = \sum_{i=1}^N  \left[ \bar{y}_i - f_i( \theta)\right]^2 s_{n,i}^{-2} \e.
\eeq 
The stationarity conditions then reduce to
\beq \label{eq:sigmastat}
	\sum_{i=1}^N  \frac{ \bar{y}_i}{s_{n,i}^2}  \, \partial_{\theta} f_i 
	= \sum_{i=1}^N  \frac{1}{s_{n,i}^2} \, f_i (\theta)  \,  \partial_{\theta} f_i \e,
\eeq
the solution of which gives us our estimate, $\hat{\theta}[y]$, for $\theta$ for a given data-set $y = (y_{i,j})_{i=1,\ldots, N; j=1,\ldots,n}$. Note, because $N$ and $n$ are finite the estimate for $\theta$ resulting from Eq.~(\ref{eq:sigmastat}) is a stochastic quantity and will vary from one experimental realization to another.

To calculate the bias of the estimator for given $n$, we need to find its expectation value for that $n$.  
We do this in two steps:  First, we let the number of data-points $N \rightarrow \infty$, while keeping $n$ and the number of fit-parameters fixed, as in an infinite experiment.    
In this limit the fit returns an estimate $\theta^*_n$ of $\theta$ that is no longer a fluctuating quantity, but may still depend on $n$.    
So, for $N=\infty$ Eq.~(\ref{eq:sigmastat}) reads
\beq \label{eq:Nbig}
	\sum_{i =1}^{\infty}  \frac{\bar{y}_i}{s_{n,i}^2} \,   \partial_{\theta } f_i (  \theta^*_n ) 
	= \sum_{i =1}^{\infty} \frac{1}{s_{n,i}^2} \,  f_i( \theta^*_n ) \,  \partial_{\theta } f_i ( \theta^*_n ) \e.
\eeq
Note, that the experimental values, $\bar{y}_i$ and $s_{n,i}$, are still fluctuating quantities described by the same statistic as before, whereas $f_i(\theta^*_n)$ and $\partial_{\theta} f_i(\theta^*_n)$ are not fluctuating because they are functions of non-fluctuating variables.  
So, when in the next step we take the expectation value of Eq.~(\ref{eq:Nbig}), only $\bar{y}_i$ and $s_{n,i}$ are affected
\beq \label{eq:limit}
	\sum_{i =1}^{\infty}  \LA \frac{\bar{y}_i}{s_{n,i}^2} \RA \,   \partial_{\theta } f_i ( \theta^*_n ) 
	= \sum_{i =1}^{\infty} \LA \frac{1}{s_{n,i}^2} \RA \,  f_i ( \theta^*_n ) \,    \partial_{\theta } f_i ( \theta^*_n ) \e,
\eeq
which is solved by
\beq \label{eq:sigmabias}
	 f_i ( \theta^*_n)  
	= \frac{ \LA \bar{y}_i / s_{n,i}^{2} \RA }{ \la  1/s_{n,i}^{2} \ra}, 
	\mbox{  for all } i=1,\ldots,\infty \e,
\eeq
if a parameter set $\theta^*_n$ exists, which solves all these many equations simultaneously.
Provided that we know the distribution of $y_i$, we can calculate the expectation values at the right-hand side of this equation and thus determine whether the fit is biased.
For a start, note that if the sample mean and the reciprocal sample variance are uncorrelated, then the 
numerator on the right-hand-side factorized to $ \la \bar{y}_i \ra \la 1 / s_{n,i}^{2} \ra$.  
Thus the right-hand-side equals $ \la \bar{y}_i \ra $, i.e. the fit is unbiased. So it is sufficient that mean and reciprocal variance are uncorrelated to ensure an unbiased fit.  
  
It turns out that it is also a necessary condition, if there is no redundancy in the parameterization of $f_i$ by $\theta$, i.e., if their relationship is one-to-one.  This is the only sensible way to parameterize a function, and easily achieved by elimination a possible redundancy through reparameterization.  Assuming this one-on-one relationship, we use that $f_i(\theta^*)  = \la \bar{y}_i \ra$ and Eq.~(\ref{eq:sigmabias}) to write
\bea \label{eq:squarebracket}
	f_i(\theta^*)  &=& \left[  \la \bar{y}_i \ra \,  
	\frac{ \LA 1 / s_{n,i}^{2} \RA }{ \la  \bar{y}_i/s_{n,i}^{2} \ra} \right] 
	\,  f_i ( \theta ^*_n) , 	\mbox{  for all } i=1,\ldots,\infty \e. \nonumber \\	
\eea
We now see  that the fit is unbiased if and only if the term in square brackets in Eq.~(\ref{eq:squarebracket}) equals unity.

In summary, a \emph{necessary and sufficient criterion for least-squares fitting to be unbiased is that the sample mean and reciprocal sample variance are uncorrelated.}
Note that `uncorrelated' is a weaker requirement than `independent' and although the latter implies the former, the reverse is not generally true.
Naturally, when the sample mean and sample variance are independent, the sample mean and reciprocal sample variance are also uncorrelated.

Notice that we nowhere used what $s_{n,i}$  is, and in fact a more general version of Eq.~(\ref{eq:squarebracket}) is
\beq \label{eq:squarebracket2}
	f_i(\theta^*)  = \left[  \la {y}_i \ra \,  
	\frac{ \LA w_i^{2} \RA }{ \la {y}_i \, w_i^{2} \ra} \right] 
	\,  f_i ( \theta ^*_n)  	
\eeq
where $w_i$ is assumed independent of $\theta$ but is otherwise unconstrained.
The criterion for an unbiased estimator is now simply that the data, $y_i$, and the squared weights, $w_i^2$, are uncorrelated.

Under some circumstances it is possible to determine the bias of the fit parameters from the bias of the fit-function given in Eq.~(\ref{eq:squarebracket2}).  If the function is invertible this is trivially the case. 
But, it is also possible if the amplitude of the fit-function is determined by just one of the fit-parameters and the term in the square brackets is independent of $i$. 
In this case, bias can be removed from the parameter estimate, $\hat{\theta}$,  by simply multiplying the amplitude-parameter by the square bracket and leaving the other parameters unchanged.  In the next section we give an example of this.

\subsubsection{Experimental averages as weights}
If the  sample standard deviation is proportional to the sample mean, $s_{n,i} \propto \bar{y}_i$ for each $i$, the mean and variance are clearly not uncorrelated and we expect the parameter estimate to be biased.  From Eq.~(\ref{eq:squarebracket}) we find  
\beq \label{eq:squarebracket3}
	f_i ( \theta^* ) = \left[ \la\bar{y}_i \ra  
	\frac{ \la \bar{y}_i^{-2} \ra}  { \la \bar{y}_i^{-1}  \ra} \right] f_i ( \theta^*_n ) \e.
\eeq

As a real-world example, consider gamma-distributed experimental data.
We already mentioned that the gamma-distribution describes all power-spectra resulting from linear dynamical equations driven by an additive noise, white or colored.
Carrying on as before, we use the standard deviation to weight the data, $w_f = 1/s_n(\Pbarexp_f)$, where $f$ is the frequency.
Since the standard deviation and the mean are proportional, see Eq.~(\ref{eq:Gammasigma}), we have $w_f \propto 1/\Pbarexp_f$. 
Using $\Pbarexp_f$'s known gamma-distribution, see Eq.~(\ref{eq:gammadist}), we can compute $\la (\Pbarexp_f)^{-q} \ra$ for $q=1,2,\ldots$ and find the value of the bracketed bias-term in Eq.~(\ref{eq:squarebracket3})
\beq
	\left[ \la \Pbarexp_f \ra  
	\frac{ \la (\Pbarexp_f)^{-2} \ra}  { \la  ( \Pbarexp_f ) ^{-1}  \ra} \right] = n / (n-2) \e,
\eeq
That is, \emph{weighted least squares fitting of this wide class of power-spectra has a built-in, frequency independent, multiplicative bias of $\frac{n-2}{n}$ for $n>2$}.
The true power spectral form can then be obtained from the least-square fit as
\beq \label{eq:Pbias}
	P_f( \theta^*) = \frac{n}{n-2} \, P_f( \theta^*_n ) 
\eeq
In this scenario, the theoretical $N \rightarrow \infty$ limit that we took in going from Eq.~(\ref{eq:sigmastat}) to Eq.~(\ref{eq:Nbig})  corresponds to letting the measurement time (or sampling frequency) become infinite while keeping the sampling frequency (or measurement time) and all other experimental factors fixed. 

In the simple case of a Lorentzian, Eq.~(\ref{eq:Lorentzian}),  we have $\theta^* = (D, f_c)$ and
\beq
	P_f( \theta^* ) \equiv  \frac{D/(2\pi^2)} {\fc^2+ f^2}=  \frac{n}{n-2} \frac{D^*_n/(2\pi^2)} {(f^*_{{\rm c}, n})^2+ f^2} \e,
\eeq
from which we see that $f^*_{{\rm c}, n}$ is unbiased.
In other words, the true value of the fit parameters can be obtained from the WLS estimates as  
\bea
	D   &=& \frac{n}{n-2} \,  D^*_n  \approx \frac{n}{n-2} \,  D\lsq \\
 	\fc  &=&  f^*_{{\rm c}, n}  \approx \fc\lsq \e,
\eea
where $D\lsq$  and $\fc\lsq$ are the stochastically fluctuating values returned by a least-square fit to a finite $N$ data-set.
This is a general feature of the power-spectra: They can be written as the product of an overall multiplicative scale-factor and a shape-function depending on $f$, where \emph{only the scale-factor is influenced by the bias}.

\subsubsection{Theoretical values as weights}

If the sample standard deviation is known to be proportional to $f_i(\theta^*)$, it seems reasonable to weight the data by the theoretical value $1/f_i$.
After inserting $w_i \propto 1/f_i$ in Eq.~(\ref{eq:chi2}), the quantity to minimize is
\beq 
	\label{eq:chi2theo}
	\chi^2 (\theta) = \sum_{i=1}^{N} \left[ y_i - f_i(\theta) \right]^2 f_i^{-2}(\theta) \e,
\eeq 
and the stationarity equations Eq.~(\ref{eq:genstat}) become
\beq
	\sum_{i=1}^N y_i \, f_i^{-2} \, \partial_{\theta} f_i =
	\sum_{i=1}^N y_i^2 \, f_i^{-3} \, \partial_{\theta} f_i  \e.
\eeq
Repeating the arguments from the previous section, the expectation value of the stationarity conditions in the limit $N \rightarrow \infty$ is
\beq
	\sum_{i=1}^{\infty} \la y_i \ra \, f_i^{-2} ( \theta^*_n )   \, \partial_{\theta}  f_i  ( \theta^*_n )  
	= \sum_{i=1}^{\infty}  \la y_i^2 \ra \,  f_i^{-3} ( \theta^*_n )  \, \partial_{\theta} f_i  ( \theta^*_n )  \e,
\eeq
which is solved by
\beq 
	 f_i ( \theta^*_n )  = \frac{\la y_i^2 \ra}{ \la y_i \ra} ,
	 \mbox{  for all } i=1,\ldots,\infty \e,
\eeq
if a parameter set $\theta^*_n$ exists, which solves all these many equations simultaneously.
The true value, $f_i(\theta^*)$, can in that case be written as
\beq \label{eq:ftrue}
	 f_i ( \theta^*)  = \left[  {\la y_i \ra ^2 }/{ \la y_i^2 \ra}  \right] f_i ( \theta^*_n )
\eeq
and $\theta^*$ can be determined from $\theta_n^*$ if $\la y_i \ra^2 / \la y_i^2 \ra$ is independent of $i$, constant, and can be absorbed in $\theta_n^*$ in a simple manner.

If we again use the gamma-distributed power spectral data as example, we see that 
\beq \label{eq:Pbiastheo}
	P_f( \theta^*) = \frac{n}{n+1} \, P_f( \theta^*_n ) 
\eeq 
i.e, we once more have a bias that scales with the number $n$ of spectra averaged over, although this time the bias is half the size and positive.

Closed-form expressions for $f^*_{{\rm c}, n}$ and $D^*_n$ are straightforward to obtain, when $P_f$ is a Lorentzian \cite{Berg-Sorensen2004} (reference \cite{Berg-Sorensen2004}  contains a  typo: $D T_{\rm msr}$ should simply read $D$). 
The expression for $f^*_{{\rm c}, n}$ given in \cite{Berg-Sorensen2004} is un-biased, whereas the expression for $D^*_n$ has its bias removed by multiplying by $n/(n+1)$. 
Note, that apart from the  $n/(n+1)$ factor on $D$, the results derived in \cite{Berg-Sorensen2004} are mathematically identical to those derived using an MLE approach in the next section.

\subsubsection{Constant values as weights (``un-weighted")}
\begin{figure}[ht]
 \includegraphics[width=8cm]{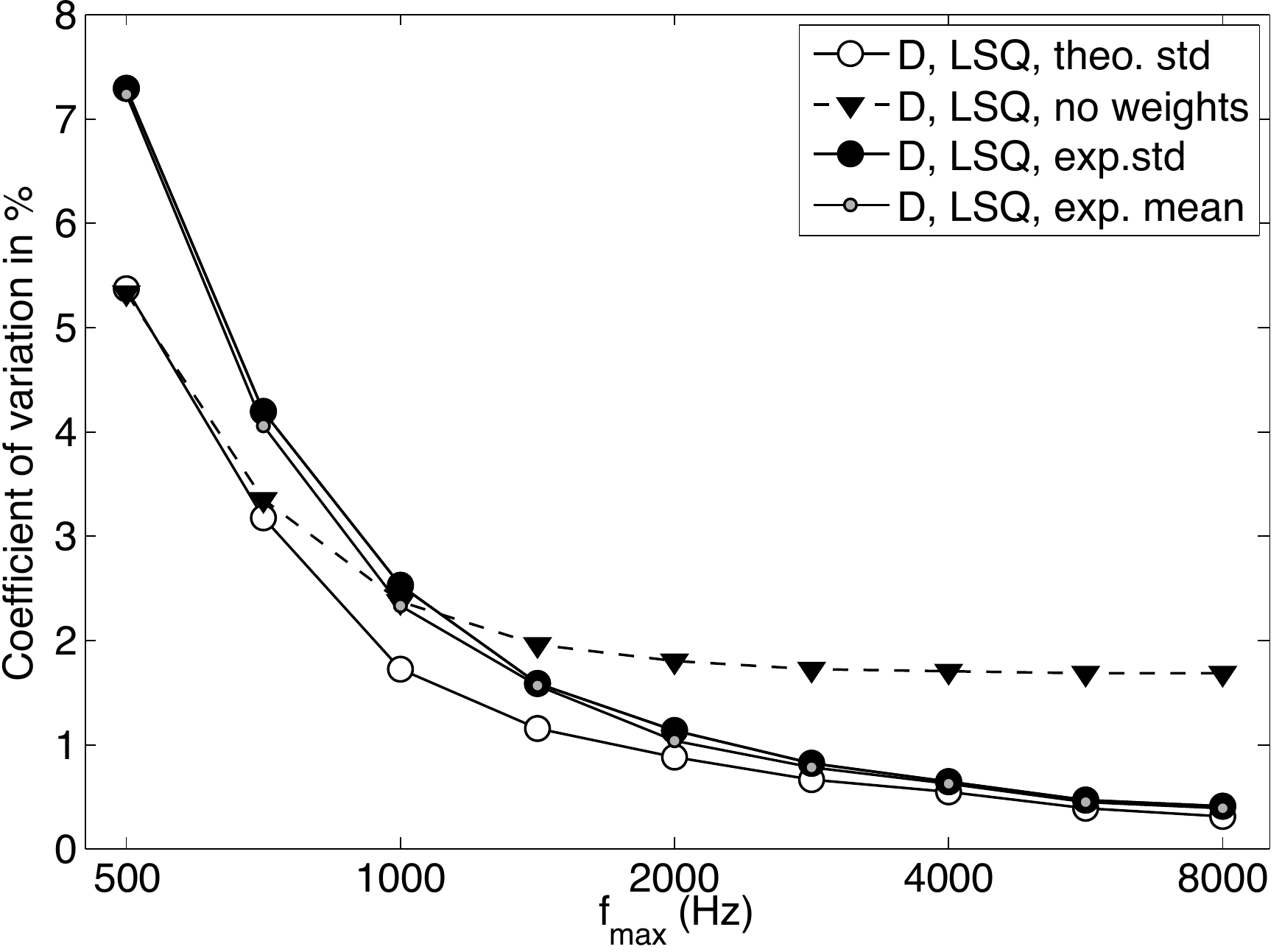}
\caption[]{\label{fig:aliasLSQ_variance}
Stochastic error on fit parameters as a function of cut-off frequency, $f_{\rm max}$, for least squares fitting of aliased Lorentzians using various weights, Eq.~(\ref{eq:chi2}).
For all fits the errors are substantial for low values of $f_{\rm max}$.
Notice the logarithmic frequency scale: For large $f_{\rm max}$, WLS returns a variance nearly an order of magnitude smaller than for OLS (fitting with constant weights).
The total number of acquired data points was held fixed at  $N=262,144$, and the fit was done to the average of $n=16$ power spectra generated with $\fc=500$\,Hz, $D=0.46\,\mu$m/s$^2$, and $\fsample=16,384$\,Hz.
Symbols show the coefficient of variation, $s_n(D)/\bar{D}$, with $D$ measured over 100 independent stochastic simulations.  
The coefficient of variation for $\fc$ show similar trends and are roughly two times larger at these settings (data not shown).
For OLS (triangles, Eq.~(\ref{eq:chi2constant})) the error is independent of $f_{\rm max}$ for large frequencies because the information there is de-emphasized by the fitting algorithm.
Using the experimental standard deviation or mean as weights (filled circles, Eq.~(\ref{eq:chi2sigma})) leads to \emph{stochastic} errors nearly as small as when using theoretical weights (empty circles, Eq.~(\ref{eq:chi2theo}))---the \emph{systematic} errors (biases) are twice as big and of the opposite sign however, see Fig.~\ref{fig:aliasLSQ_error}.
}
\end{figure}

\begin{figure}[ht]
 \includegraphics[width=8cm]{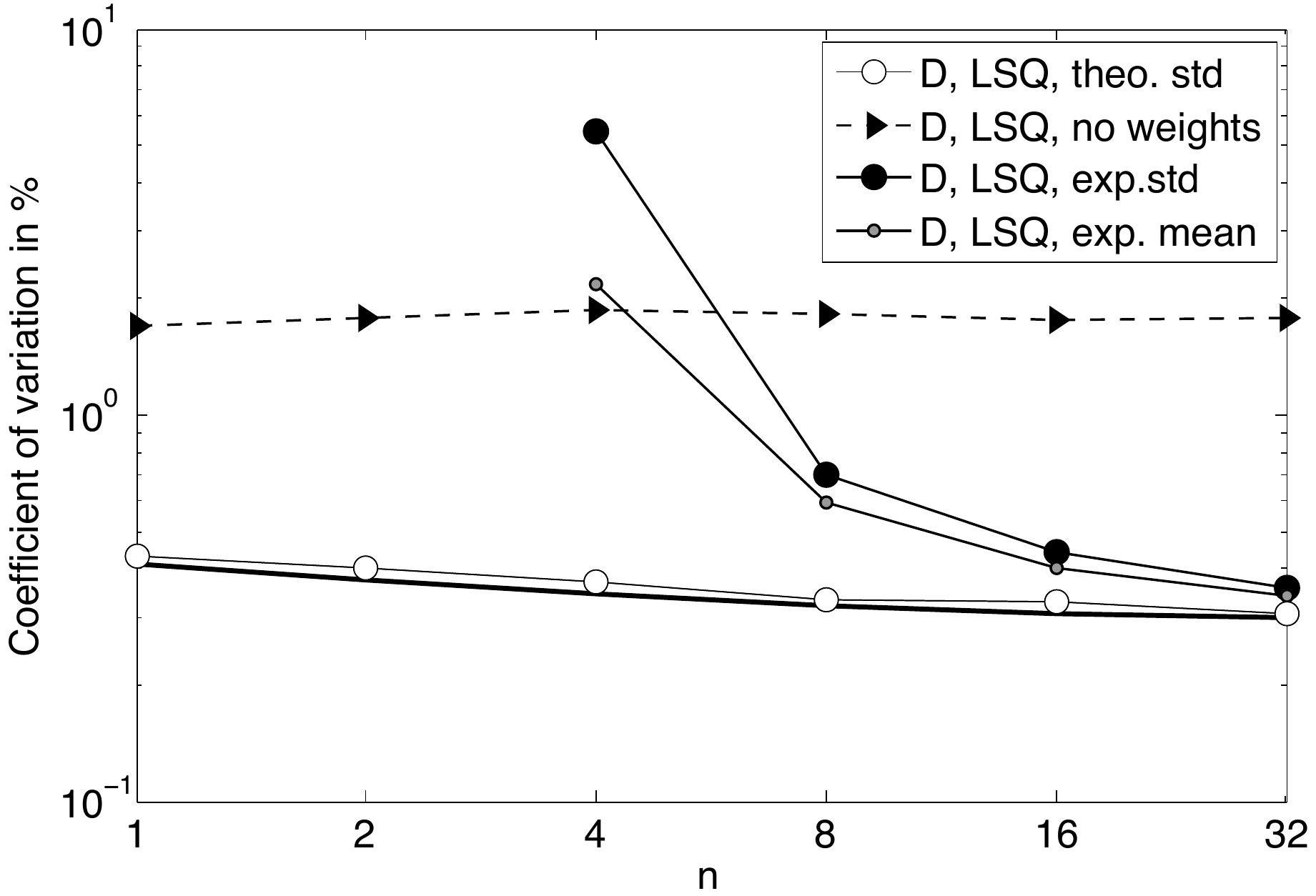}
\caption[]{\label{fig:aliasLSQ_variance_n}
Stochastic error on fit parameters as a function of number of power spectra averaged over, $n$, for least squares fitting of aliased Lorentzians using various weights, Eq.~(\ref{eq:chi2}):
Constant weights (triangles); experimental standard deviation (black circles); experimental average (grey circles); theoretical weights (white circles). Thick black line through white circles show the theoretical prediction Eq.~(\ref{eq:varD}).
The number of acquired data points was held fixed at  $N=262,144$, and the fit was done to all the data, $f_{\rm max} = \fNyq$, for power spectra generated with $\fc=500$\,Hz and $D=0.46\,\mu$m/s$^2$.
Symbols show the coefficient of variation, $s_n(D)/ \bar{D}$, with $D$ measured over 1,000 independent stochastic simulations.
Using theoretical weights clearly outperforms all the other weighing schemes for $n < 32$;
for $n>32$, the experimental weighing scheme performs comparably well---use of the experimental standard deviation as weight leads to slightly larger variation than use of the experimental average, as expected, and both show more variation than results obtained with theoretical weights.  
Results obtained with constant weights---sometimes referred to as ``no weights'' or ``un-weighted''---show roughly ten times larger variation than does results obtained with theoretical weights, for all $n$. }
\end{figure}

When all weights are assumed to be equal, the function to minimize,
\beq  \label{eq:chi2constant}
	\chi^2 (\theta) = \sum_{i=1}^N \left[ y_i - f_i (\theta) \right]^2 \e,
\eeq 
has stationarity equations which,
for $N\rightarrow\infty$ gives
\beq
	\sum_{i=1}^{\infty} f_i(\theta_n^*) \, \partial_{\theta} f_i(\theta_n^*) 
	= \sum_{i=1}^{\infty} y_i \, \partial_{\theta} f_i(\theta_n^*) \e.
\eeq
Taking the expectation value on both sides, we have 
\beq
	\sum_{i=1}^{\infty} f_i(\theta_n^*) \, \partial_{\theta} f_i(\theta_n^*) 
	= \sum_{i=1}^{\infty} \langle y_i \rangle \, \partial_{\theta} f_i(\theta_n^*) \e.
\eeq
This equation may have the solution
\beq
	 f_i (\theta^*_n)= \la y_i \ra = f_i(\theta^*)\e.
\eeq
In other words, this estimator is unbiased.
However, it is not a precise estimator: The stochastic errors on values it returns  tend to be larger than those obtained with a weighted fit.  Figures~\ref{fig:aliasLSQ_variance} and \ref{fig:aliasLSQ_variance_n} illustrate this for the case of fitting an aliased Lorentzian to power spectra from optical tweezers.

\begin{figure}[ht]
 \includegraphics[width=8cm]{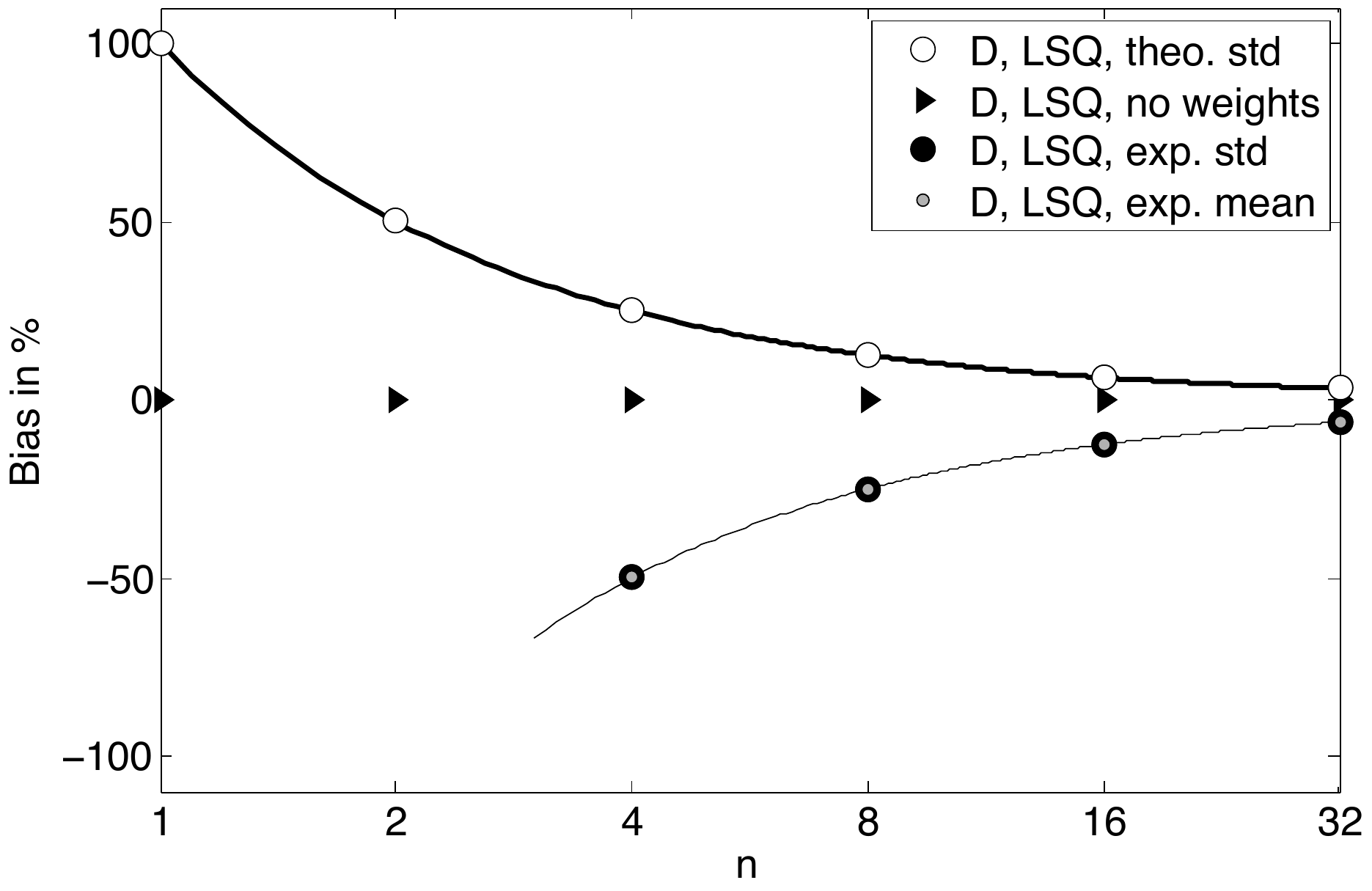}
\caption[]{\label{fig:aliasLSQ_error}
Systematic error (bias) on fit parameters as a function of number of power spectra averaged over, $n$,  for least squares fitting of aliased Lorentzians using various weights,  Eq.~(\ref{eq:chi2}).
The corner frequency is unbiased (data not shown), independent of $n$, whereas the diffusion coefficient shows a strong weighing-scheme dependent bias.
The sampling frequency was held fixed at   $\fsample=16,384$\,Hz, and the fit was done to the averaged  power spectra generated with  $N=262,144$ ($\Tmsr = 16$\,s), $\fc=500$\,Hz, and $D=0.46\,\mu$m/s$^2$.
Symbols show the signed error on the mean of $D$ measured over 1,000 independent stochastic simulations.
For least squares fitting with constant weights (triangles) the bias on $D$ is zero.
Least squares fitting with experimental weights (black circles: standard deviation; grey circles: average value) has a systematic error on $D$ with $-2/n$ dependence (thin black line) for $n>2$.
Theoretical weights (white circles) lead to an $1/n$-dependence (thick black line) in the systematic error on $D$.}
\end{figure}

\subsubsection{Iterated values as weights}
When weights are determined iteratively, a fit is performed and the value returned for $\theta$ is used in the weight of a new fit and so on, until some convergence criterion has been met. 
The function to minimize with respect to $\theta$ for given $ \theta ^{\rm (iter)}$  reads
\beq 
	\chi^2 (\theta) = \sum_{i=1}^N \left[ y_i - f_i ( \theta) \right]^2  w_i^2( \theta ^{\rm (iter)} ) \e.
\eeq 
It has the stationarity equations
\beq
	\sum_{i=1}^N  f_i \,  w_i^2( \theta ^{\rm (iter)} ) \, \partial_{\theta} f_i 
	=  \sum_{i=1}^N  y_i \,  w_i^2( \theta ^{\rm (iter)} ) \, \partial_{\theta} f_i \e.
\eeq
We now use the same arguments as above together with the observation that at the fixed-point solution to this iterative scheme, $\theta^{\rm (iter)} = \theta_n$, and both equal $\theta_n^*$ in the limit $N\rightarrow\infty$.
Taking the expectation value of the equation after taking the limit $N\rightarrow\infty$, we see that $\theta_n^*$ may satisfy
\beq
 	f_i (\theta^*_n)  = \la y_i \ra  = f_i(\theta^*) \e,
\eeq
in which case this estimator is unbiased.  
The estimation scheme just described is also known as ``iteratively reweighted least squares" and convergence is not guaranteed.

\subsubsection{Error-bars}
Error-bars on the parameter estimates $\hat{\theta}$ for $\theta^*$, can be calculated 
from the error-bars on the estimator $\hat{\theta}_n$ for $\theta_n^*$
and the simple relationship in terms of $n$ between $\theta^*$ and $\theta_n^*$,  by propagation of errors (see Eq.~(\ref{eq:differential})), provided the latter exists. 
The error bars on $\hat{\theta}_n$ can be those returned by a least-squares fitting routine used to determine $\hat{\theta}_n$, or they may be know theoretically.  In that connection, note that the error-bars given in \cite{Berg-Sorensen2004} for WLS with theoretical weights are correct only in the limit $n=\infty$, but by replacing $D$ with $n/(n+1)D$ they are correct for all $n$. 
 
When $\hat{\theta}$ is found from a $\hat{\theta}_n$ which itself is found by minimizing $\chi^2$, the precision with which  $\chi^2$ is known determines the precision of the estimate $\hat{\theta}$. 
Because the various weights in $\chi^2$ have different stochastic fluctuations,  we expect that theoretical weights, Eq.~(\ref{eq:chi2theo}), will give smaller variation in $\hat{\theta}$ than experimental weights, Eq.~(\ref{eq:chi2sigma}).  
A small calculation shows that the variance of the stationarity equations, Eq.~(\ref{eq:sigmastat}), around zero includes a term $\la \Pbarexpto{-4}_f \ra \propto \{(n-1)(n-2)(n-3)(n-4)\}^{-1}$, i.e., the error-bars are expected to be large for  $n\sim 4$, if experimental weights are used.
When theoretical weights are used, the error-bars are small and well-defined for all $n$. 
These results are verified by example, see Fig.~\ref{fig:aliasLSQ_variance_n}.

\subsection{Results for power spectra}
We verified the above theoretical results by Monte-Carlo simulations of the OT and AFM power-spectra and found perfect agreement.
Below, we list the specific results.

For the diffusion coefficient $D$, we found that the outcome of least-squares fitting, $\Dlsq$,
depends on the number, $n$, of power spectra averaged over in the following manner for the aliased (see Fig.~\ref{fig:aliasLSQ_error}) and non-aliased Lorentzians, as well as the aliased and non-aliased AFM expressions (data not shown):
\begin{enumerate}
\item $\Dlsq  = D (n-2)/n$, $n>2$; if data-points are weighted with their experimental standard-deviation or mean, $w_f = 1/s_n(\Pbarexp_f)$ or $w_f = \sqrt{n}/\Pbarexp_f$.  This is the weighting choice with the largest bias and with large stochastic errors for $n\leq4$. Here, $D$ is underestimated.
\item $\Dlsq  = D (n+1)/n$; if data-points are weighted using the theoretical expectation value for the standard-deviation, $w_f = \sqrt{n}/P_f$.  This is the least squares fitting of \cite{Berg-Sorensen2004}; a hybrid of MLE and LSQ, analytically tractable and thus very robust.  It also has the smallest stochastic errors. Here, $D\lsq$ overestimates $D$.  
\item $\Dlsq  = D $; if the weights are updated iteratively to the fitted value, $w_f=1/P_f^{\rm iter}$. This is a practically correct but computationally intensive and somewhat unstable approach; the initial guess
must be close to the true value, or the algorithm will not converge. 
\item $\Dlsq  = D $; if all data-points are given the same weight, $w_f=$ constant. This method de-emphasizes information available at higher frequencies because power spectral values have constant \emph{relative} error, hence rapidly decreasing \emph{absolute} error beyond $\fc$ or $f_0$, yet are treated as having \emph{constant} absolute error. 
Consequently, this estimator has low precision, typically with two to ten times larger stochastic errors than the  Cases 1--3 above.  This is OLS fitting applied far outside its range of validity.
\end{enumerate}
Figures \ref{fig:aliasLSQ_variance},~\ref{fig:aliasLSQ_variance_n},~and~\ref{fig:aliasLSQ_error} show examples of the size of these stochastic errors and biases, respectively, for fits of the aliased Lorentzian to data.

We found that the expectation values of the fitted values for $\fc$, $f_0$, and $Q$ were independent of weighing scheme and un-biased, whereas the standard deviations varied by up to an order of magnitude depending on weighing scheme applied.  Furthermore, fitting of non-aliased PSDs to aliased data introduces large systematic errors for the trivial reason that they fail to capture the shape of the PSD near the Nyquist frequency (data not shown).

\section{Maximum Likelihood Estimation of fit parameters}
\label{app:maxlikelihood}

We now proceed to determine the fit-parameters by the method of maximum likelihood estimation.
We also derive closed-form expressions for the fit-parameters in terms of the experimental data values. They should be useful for speedy online calibration.

Both power spectra, given in Eqs.~(\ref{eq:Lorentzian}) and (\ref{eq:AFM}), are consequences of a linear, time-invariant dynamics driven by a white noise, Eq.~(\ref{eq:NewtonMain}).
They are consequently of the form
\beq \label{eq:Pabc}
	P_f = \frac{1}{a + bf^2 + cf^4} \e,
\eeq
with $c=0$ in the case of the Lorentzian, and $(a, b)$ or $(a, b, c)$ parameters to be fitted.
This simple form, combined with the simple statistical properties of the experimental power spectral values, makes rigorous MLE of the parameters a straightforward numerical optimization problem as shown below.

When fitting, we fit only to the positive-frequency part of the power spectrum, or to a subset of it.
So here we considered only that part of the spectrum.
Since $\Pbarexp_f$ and $\Pbarexp_{f'}$ are uncorrelated for $ f\neq f'$, the probability density for the experimental spectrum $\Pbarexp_f$, given its expectation value $P_f$, is
\beq \label{eq:pPexp}
	p(\Pbarexp|P) =\prod_f  p_n(\Pbarexp_f;P_f)
	\enspace,
\end{equation}
where $p_n$ is given in Eq.~(\ref{eq:gammadist}).
Thus {\em Maximum Likelihood estimation} of the theory's parameter values consists in choosing these parameters so they maximize $p(\Pbarexp|P)$ for given $\Pbarexp$, or, equivalently, minimize the negative logarithm
\beq
	{\cal F'} \equiv n \sum_f  \left( \Pbarexp_f/P_f + \ln P_f \right) + \mbox{constant} \enspace.
	\label{eq:costfunction2}
\end{equation}
where
\beq
	\mbox{constant} = \ln \Gamma(n) - n\ln n - (n-1) \ln \Pbarexp_f 
\eeq
is a constant with respect to $(a,b,c)$ so it can be ignored when minimizing.
We are then left with the task of finding the values of $(a,b,c)$ that minimize the cost function
\beq
	{\cal F}(a,b,c) \equiv  \sum_f  \left( \Pbarexp_f/P_f + \ln P_f \right) \e,
	\label{eq:costfunction}
\eeq
which is an uncomplicated optimization problem
that can be solved numerically with standard programs.
Good starting values are given in Eqs.~(\ref{eq:fc}) and (\ref{eq:D}).

\subsection{From non-linear to linear stationarity equations via a simple trick}
\label{app:simpletrick}
Although we gave the solution to the full MLE problem implicitly above, as the minimum of $\mathcal{F}$ Eq.~(\ref{eq:costfunction}), we can speed up the fitting process substantially
(numerical optimization can be slow when the data-sets are large) and gain more insight into the problem by taking a few more analytical steps before turning to numerics.

\subsubsection{Results for Optical Tweezers}
For $P_f$ given in Eq.~(\ref{eq:Lorentzian}), written as in Eq.~(\ref{eq:Pabc}) with $c=0$, ${\cal F}$ in Eq.~(\ref{eq:costfunction}) reads
\beq
	{\cal F}(a,b) =  \sum_f  \left((a+bf^2) \Pbarexp_f   - \ln(a+bf^2) \right) \enspace.
\eeq
It is minimized with respect to $a$ and $b$ when these parameters satisfy the stationarity condition
\bea
	\sum_f  \Pbarexp_f   &=& \sum_f P_f = \sum_f \frac{1}{a + b f^2} \nonumber \\
	\sum_f  f^2 \Pbarexp_f   &=& \sum_f  f^2 P_f = \sum_f \frac{f^2}{a + b f^2}\enspace.
	\label{eq:nonlin}
\eea
These are non-linear equations for $a$ and $b$.
However, from Eq.~(\ref{eq:qmoment}) we know that we can write
\beq \label{eq:trueid}
	P_f= \frac{n\langle \Pbarexpto{2}_f\rangle}{(n+1) P_f}  
	= \frac{n\langle \Pbarexpto{2}_f\rangle}{(n+1)} (a+bf^2) \enspace.
\eeq
Substituting for $P_f$ in Eq.~(\ref{eq:nonlin}) we find
\bea
	\sum_f  \Pbarexp_f   &=& \frac{n}{n+1} \sum_f \la \Pbarexpto{2} \ra (a + b f^2) \nonumber \\
	\sum_f  f^2 \Pbarexp_f   &=&  \frac{n}{n+1} \sum_f f^2\la \Pbarexpto{2} \ra (a + b f^2) \enspace,
	\label{eq:lin}
\eea
which is linear in $a$ and $b$.

Before solving for $a$ and $b$ we introduce the statistic
\beq \label{eq:Spq}
	S_{p,q} \equiv  \frac{1}{K}\sum_f f^{2p} \Pbarexpq_f  \enspace,
\eeq
with $K$ the number of terms in the sum, and the statistic obeys $\lim_{K \rightarrow \infty} S_{p,q} = \la S_{p,q} \ra$.

The sums should only include those frequencies the user deems relevant,
i.e., frequencies corresponding to mechanical or electronic resonances can be excluded,
and high/low frequency cut-offs can be applied.
That is, the statistics can be trimmed iteratively if required:
Power spectral values too far from a fit can be identified and excluded from the sums, after which a new fit is found, et cetera until a steady state is reached and all power spectral values satisfy the user-defined acceptance criterion.
When no frequencies are excluded, $nK=N=\Tmsr \fsample$, the total number of data acquired.
In this latter case, one fits the power spectrum all the way out to the Nyquist frequency,
and aliasing should be taken into account (see appendices);
unless aliasing is eliminated by over-sampling data acquisition electronics.

With this notation, Eq.~(\ref{eq:lin})  can be written in matrix form
\beq
	\left(\begin{array}{cc}
	\la S_{0,2} \ra & \la S_{1,2} \ra   \\
	\la S_{1,2} \ra & \la S_{2,2} \ra
	\end{array}\right)
	\left(\begin{array}{c}
	a \\
	b
	\end{array}\right)
	= \frac{n+1}{n}
	\left(\begin{array}{c}
	S_{0,1}  \\
	S_{1,1}
\end{array}\right) \e,
\eeq
with solution
\beq \label{eq:solution} 
	\left(\begin{array}{c}
	a \\
	b
	\end{array}\right)
	= \frac{1+1/n}{ \la S_{0,2} \ra \la S_{2,2}\ra - \la S_{1,2} \ra ^2 }
	\left(\begin{array}{c}
	S_{0,1} \la S_{2,2} \ra -S_{1,1} \la S_{1,2} \ra  \\
	S_{1,1} \la S_{0,2} \ra  -S_{0,1} \la S_{1,2} \ra
	\end{array}\right) \e,
\eeq
In this expression, we know $S_{p,q}$ from the experiment, but not $\la S_{p,q} \ra$.  
However, we can always write $S_{p,q} = \la S_{p,q}\ra + \delta S_{p,q}$ and substitute for $\la S_{p,q}\ra$ in Eq.~(\ref{eq:solution}).  
About $\delta S_{p,q}$ it is easy to show that $ \la \delta S_{p,q} \ra = 0$ and
\bea
	 \la (\delta S_{p,q})^2 \ra 
	&=& \frac{1}{K^2} \left[ \frac{ \Gamma(n+2q)}{n^{2q} \Gamma(n)} -
	 \left( \frac{ \Gamma(n+q)}{n^{q} \Gamma(n)} \right)^2 \right] 
	 \sum_f f^{4p} P_f^{2q} \nonumber \\
	 &=& \frac{1}{K}\, [\mbox{weak } n\mbox{-dep.}] \, \tilde{S}_{2p,2q} \e,
\eea
where $\tilde{S}$ is defined as in Eq.~(\ref{eq:Spq}) except for $P_f=\la \Pbarexp_f \ra$ replacing $\Pbarexp_f$.
Above, $q=2$ in the relevant expressions and therefore $\tilde{S}_{p,q}$ scales as $1/K$ for $p=0,1$ and is independent of $K$ for $p=2$. 
Thus, apart from some $n$-dependence, the variance of $S_{p,q}$ scales as $1/K^2$ or $1/K$ both of which are very small numbers in a typical experiment, where $K = N/n$ is of order  $10^4$--$10^6$.
It is therefore an excellent approximation to replace $\la S_{p,q} \ra$ with $S_{p,q}$ in Eq.~(\ref{eq:solution})
\beq \label{eq:appsolution} 
	\left(\begin{array}{c}
	a \\
	b
	\end{array}\right)
	\approx
	\left(\begin{array}{c}
	\atrick \\
	\btrick
	\end{array}\right)
	\equiv \frac{1+1/n}{  S_{0,2}   S_{2,2} -  S_{1,2}  ^2 }
	\left(\begin{array}{c}
	S_{0,1}  S_{2,2}  -S_{1,1}  S_{1,2}   \\
	S_{1,1}  S_{0,2}   -S_{0,1}  S_{1,2} 
	\end{array}\right) \e,
\eeq
where now $(\atrick,\btrick)$ are our estimates of $(a,b)$.
By comparing Eqs.~(\ref{eq:Lorentzian}) and (\ref{eq:Pabc}) we then have
\bea
	\fc &=& \left(\frac{a}{b}\right)^{1/2} \approx
	\left(\frac{\atrick}{\btrick}\right)^{1/2} =
	\left(\frac{ S_{0,1} S_{2,2}-S_{1,1}S_{1,2}}{S_{1,1}S_{0,2}-S_{0,1}S_{1,2}}\right)^{1/2} \label{eq:fc} \\
	D &=& \frac{2\pi^2}{b} \approx \frac{2\pi^2}{\btrick} =
	\frac{2\pi^2 n}{n+1}  \frac{ S_{0,2} S_{2,2}-S_{1,2}^2}{S_{1,1}S_{0,2}-S_{0,1}S_{1,2}} \label{eq:D} \e.
\eea

\subsubsection{Results for AFM cantilevers}
Identical reasoning leads, in the case of $c\neq 0$, to the three coupled linear equations for $a$, $b$, and $c$ 
\beq \label{eq:Svs3}
	\underbrace{
	\left(\begin{array}{ccc}
	S_{0,2} & S_{1,2}  & S_{2,2} \\
	S_{1,2} & S_{2,2}  & S_{3,2} \\
	S_{2,2} & S_{3,2}  & S_{4,2}
	\end{array}\right)
	}_{\mathbf{S}}
	\underbrace{
	\left(\begin{array}{c}
	a \\
	b \\
	c
	\end{array}\right)
	}_{\vec{v}}
	\approx (1+1/n)
	\underbrace{
	\left(\begin{array}{c}
	S_{0,1}  \\
	S_{1,1} \\
	S_{2,1}
	\end{array}\right)
	}_{\vec{s}} \e,
\eeq
which are inverted to give
\beq  \label{eq:abc}
	\left(\begin{array}{c}
	a \\
	b \\
	c
	\end{array}\right)
	\approx
	\left(\begin{array}{c}
	\atrick \\
	\btrick \\
	\ctrick
	\end{array}\right)
	\equiv \frac{n+1}{n} \mathbf{S}^{-1} 	
	\left(\begin{array}{c}
	S_{0,1}  \\
	S_{1,1} \\
	S_{2,1}
	\end{array}\right) \e.
\eeq

Comparing Eqs.~(\ref{eq:AFM}) and (\ref{eq:Pabc}) then gives
\bea
	 f_0 &=& \left(\frac{a}{c}\right)^{1/4} \approx \left(\frac{\atrick}{\ctrick}\right)^{1/4}  \label{eq:f0}\\
	 D &=& \frac{2\pi^2}{b+2(ac)^{1/2}}\approx \frac{2\pi^2}{\btrick+2(\atrick \ctrick)^{1/2}}  \label{eq:DAFM} \\
	\left( \frac{2\pi m}{\gamma}\right)^2 &=& \frac{c}{b+2(ac)^{1/2}} \approx \frac{\ctrick}{\btrick+2(\atrick \ctrick)^{1/2}}  \label{eq:G} \enspace.
\eea

For completeness and ease of implementation we give the inversion formulas:
$\mathbf{S}^{-1} = \mathbf{C}/\mathcal{D}$,
\bea
	\mathcal{D} &=& S_{0,2}S_{2,2}S_{4,2}-S_{0,2}S_{3,2}^2 \\
	&-& S_{1,2}^2S_{4,2}+2S_{1,2}S_{2,2}S_{3,2}-S_{2,2}^3 \e, \nonumber \\
	\nonumber \\
	\mathbf{C} &=&
	\left(\begin{array}{ccc}
	C_{0} & C_{1}  & C_{2} \\
	C_{1} & C_{5}  & C_{3} \\
	C_{2} & C_{3}  & C_{4}
	\end{array}\right) \e,
\eea
and
\bea \label{eq:Cs}
	C_0 &=& S_{2,2}S_{4,2} - S_{3,2}^2   \nonumber  \\
	C_1 &=& S_{2,2}S_{3,2} - S_{1,2}S_{4,2}  \nonumber\\
	C_2 &=& S_{1,2}S_{3,2} - S_{2,2}^2  \\
	C_3 &=& S_{1,2}S_{2,2}-  S_{0,2}S_{3,2} \nonumber \\
	C_4 &=& S_{0,2}S_{2,2} - S_{1,2}^2   \nonumber \\
	C_5 &=& S_{0,2}S_{4,2} - S_{2,2}^2  \enspace.  \nonumber
\eea
Of course, it is also possible to numerically invert Eq.~(\ref{eq:Svs3}) and insert the resulting values for $\atrick,\btrick,\ctrick$ in Eqs.~(\ref{eq:fc},\ref{eq:D},\ref{eq:f0},\ref{eq:DAFM},\ref{eq:G}).  The above results however, should resolve potential issues with numerical stability arising from the inversion of near-singular matrices.

\subsection{Examples of Fitting the power spectra}

\begin{figure}[ht]
 \includegraphics[width=8cm]{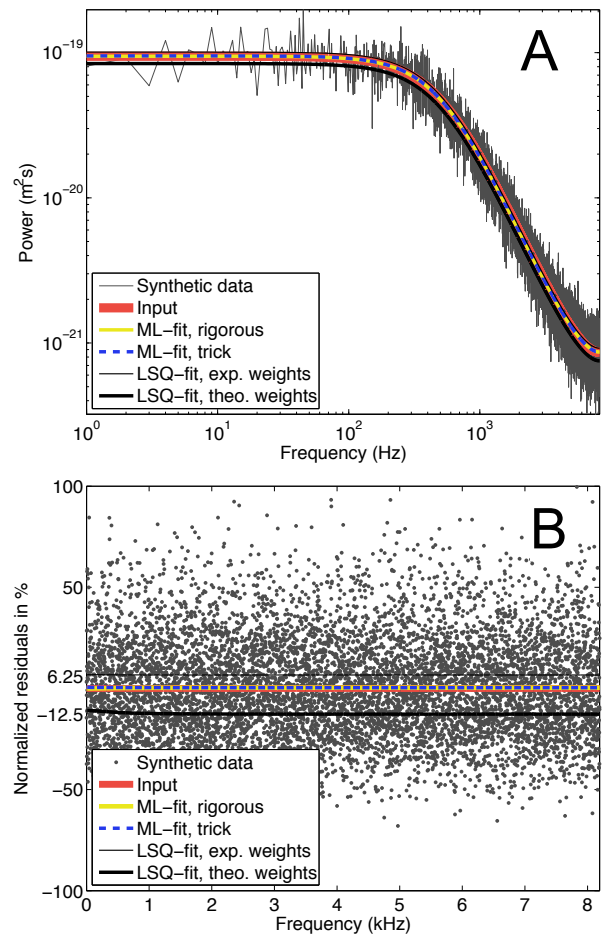}
\caption[]{\label{fig:psd1}
Various fits of aliased Lorentzian to synthetic power spectrum for microsphere in optical trap.
Data are artificial, with known parameter values $\fc=500$\,Hz and $D=0.46\,\mu$m$^2$/s, $\fsample = 16,384$\,Hz, and $\Tmsr = 16$\,s.
\textbf{A:} Data and aliased Lorentzians. PSD values (grey line) are averages over 16 spectra. Note, how the three Lorentzians corresponding to, respectively,  the exactly known values of $\fc$ and $D$  (red line),  rigorous numerical ML-fit (yellow line), and analytical ML-fit using simplifying trick (dashed blue line) all plot on top of each other.  In contrast, the Lorentzian from numerical least-squares fits with experimental or theoretical weights (thick and thin black lines, respectively) are offset by -12.5\% and +6.25\% due to their systematic errors on $D$.
\textbf{B:}  Residual plots, i.e., same data and fits as shown in Panel A, but divided by the known true expectation value, then unity subtracted.  The agreement between data and the ML-fits is practically perfect whereas the least-squares fits clearly under/over-estimate the PSD.}
\end{figure}

Figure~\ref{fig:psd1}A shows the average of $n=16$ power spectra for the Brownian motion of a micro-sphere in an optical trap.
The data are synthetic, computer generated; see \cite{Berg-Sorensen2004}.
Thus, the values for $\fc$ and $D$ are known exactly,
and can be used as benchmarks for results obtained by fitting,
with no worry about any of the complicating circumstances that can affect data in the real world.
The red line is the expectation value of the aliased Lorentzian,
taking as input the exactly known values of $\fc$ and $D$.
The yellow line is the aliased Lorentzian corresponding to
the stochastically realized parameter values for $\fc$ and $D$,
determined by rigorous MLE of $(a,b)$ with no use of our simplifying trick.
The dashed blue line is the result of MLE,  with the use of our simplifying trick.
All of these three lines plot virtually on top of each other.
Finally, the black lines shows the result of least squares fitting of an aliased Lorentzian to the data, using as weights: (i) The standard deviations on the $n = 16$ spectra with a resulting $-2/n = -12.5\%$ systematic error on $D$, see Eq.~(\ref{eq:Pbias}); and (ii) Theoretical weights, resulting in a $1/n = 6.25\%$ systematic error on $D$, see Eq.~(\ref{eq:Pbiastheo}).

\begin{figure}[ht]
 \includegraphics[width=8cm]{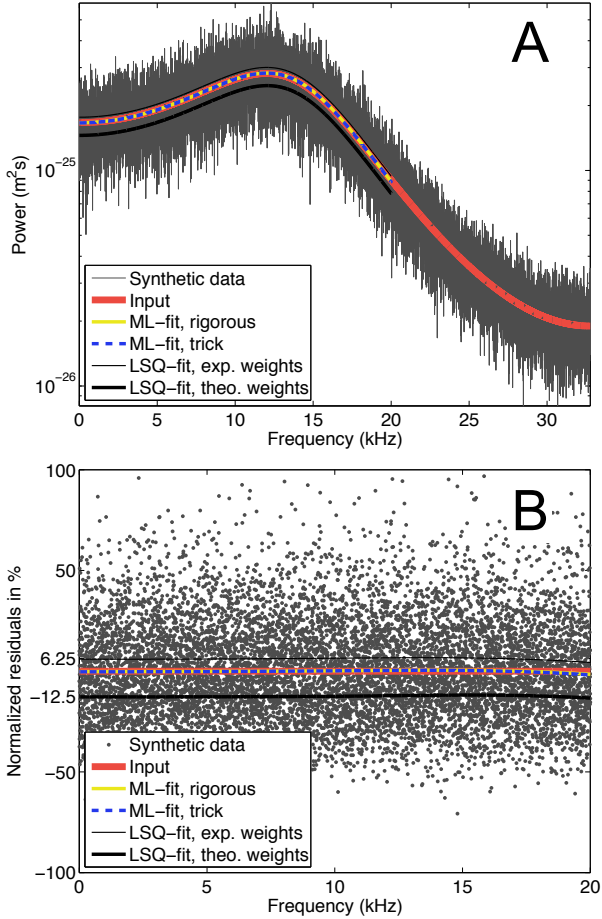}
\caption[]{\label{fig:psd2}
Various fits of Eq.~(\ref{eq:AFM}) to the average of 16 power spectra for an AFM cantilever.
Data are synthetic with known parameter values $f_0=15$\,kHz, $D=0.0010\,\mu$m$^2$/s, and $(\frac{2\pi m}{\gamma})^2=6169\,\mu$s$^2$, $\fsample = 65,536$\,Hz, and $\Tmsr = 8$\,s.
\textbf{A}: Data and fits. Note, how the three curves corresponding to, respectively,  the exactly known parameters  (red line),  rigorous numerical ML-fit (yellow line), and analytical ML-fit using simplifying trick (dashed blue line) all plot on top of each other.  In contrast, the result of a numerical least-squares fits with experimental or theoretical weights (thick and thin black lines, respectively) are offset by -12.5\% and +6.25\% due to their systematic errors on $D$.  In all the fits, only frequencies below 20\,kHz were included.
\textbf{B}: Residual plots, i.e., same as shown in Panel A, but divided by the exactly known expectation value, then unity subtracted.  At the highest frequencies, where the non-aliased fits start to deviate from the aliased PSD, a slight departure from a straight line is seen in all the fits.}
\end{figure}

Note, that for clarity of presentation we discuss the non-aliased cases in the main-text and give closed-form aliased results in Appendices~\ref{app:aliasedAFM}~and~\ref{app:alias}.
Numerical tests of the theory are done using the aliased theory in order to utilize all the available data: When fitting a non-aliased expression to aliased data it is necessary to introduce a cut-off frequency $f_{\rm max} \ll \fNyq$ and discard all data above it.

Figure~\ref{fig:psd1}B shows the same data and fits in a normalized \emph{residual plot},
i.e., data or fits minus the true expected value (the residue) divided by the true expected value.
Thus, deviation from zero in this plot shows by how much data and fits
differ from the true expected value, measured in units of the true expected value.
The data should scatter about 0 with standard deviation $n^{-1/2}=16^{-1/2}=25$\%, whereas the fits should simply trace zero for all frequencies.
Here, the true expected value is known because we use synthetic data; we normalize by it to show the results of all fits in a single figure.
In an experiment, the true expected value is not known and the fitted value is used instead---when a bias is present the residues will consequently differ from zero in a systematic manner.

This bias always hides in the scatter of the data in a figure like Fig.~\ref{fig:psd1}A, because $n^{-1} \leq n^{-1/2}$
for $n=1,2,\ldots$.
But, the bias is substantial for small to intermediate values of  $n$ and will reveal itself in a plot like Fig.~\ref{fig:psd1}B.

Figure~\ref{fig:psd2} is similar to Fig.~\ref{fig:psd1}, except its power spectrum describes the confined Brownian motion of a massive particle, e.g.\ an AFM cantilever in air, and the fits do not take aliasing into account.
The data are synthetic, computer generated; see Appendix~\ref{app:EOU-MC}.

\subsection{Error-bars on fit parameters}
\label{sec:errorbars}
Because we have derived closed-form expressions for the fit parameters we can also calculate expressions for the expected error-bars on the fit parameters.
We do that and compare to the theoretical limit for how small the error-bars can be.

Quite generally, irrespective of whether we study the aliased or non-aliased Lorentzian or the AFM PSD, the error-bars on the fit-parameters are found by propagating the errors on $a$, $b$, and $c$ using the generic formula for a function $z$ (calculating the differential $\Delta z$ and squaring it):
\bea \label{eq:differential}
	\lefteqn{\sigma^2(z[a,b,c]) = \la (\Delta z)^2 \ra = }  \\
 	&&  \left( \partial_a z \right)^2 \langle (\Delta a)^2 \rangle	+ \left( \partial_b z \right)^2 \langle (\Delta b)^2 \rangle
	+ \left( \partial_c z \right)^2 \langle (\Delta c)^2 \rangle  \nonumber\\
	&+& 2\,\partial_a z \, \partial_b z \la \Delta a \Delta b \ra
	+ 2\,\partial_a z \, \partial_c z \la \Delta a \Delta c \ra
	+ 2\,\partial_b z \, \partial_c z \la \Delta b \Delta c \ra
	\nonumber \enspace,
\eea
with $\partial_a z = \partial z / \partial a$, and similarly for $b$ and $c$;
$z=\fc,f_0,D,\ldots$,
and $\langle \Delta a \Delta b \rangle$  etc.\ are elements of the covariance matrix
\beq \label{eq:covariance}
	\mbox{cov}(a,b,c) \equiv
	\left( \begin{array}{ccc}
	\la(\Delta a)^2\ra & \la\Delta a \Delta b\ra & \la\Delta a \Delta c\ra \\
	\la\Delta a \Delta b \ra & \la(\Delta b)^2\ra  & \la\Delta b \Delta c\ra \\
	\la\Delta a \Delta c\ra & \la\Delta b \Delta c\ra & \la(\Delta c)^2\ra
	\end{array} \right) \e.
\eeq
This covariance matrix can be written in terms of the expectation values for the statistics, see Appendix~\ref{app:covariance},
\bea \label{eq:covN}
	\mbox{cov}(a,b,c) &\approx&  \mbox{cov}(\atrick,\btrick,\ctrick) = \frac{1}{N} \frac{n+3}{n}  \la \mathbf{S} \ra ^{-1}
\eea
with
\bea	
	\la \mathbf{S} \ra &=&  \frac{n+1}{n} \,  \tilde{\mathbf{S}} \e,
\eea
where $\tilde{\mathbf{S}}$ is a matrix of the same form as $\mathbf{S}$ in Eq.~(\ref{eq:Svs3}),
but with expectation values $P_f = \la \Pbarexp\ra$ replacing the experimental values $\Pbarexp$ in the statistics Eq.~(\ref{eq:Spq}), see Eq.~(\ref{eq:Stilde}).

\begin{figure}[ht]
 \includegraphics[width=8cm]{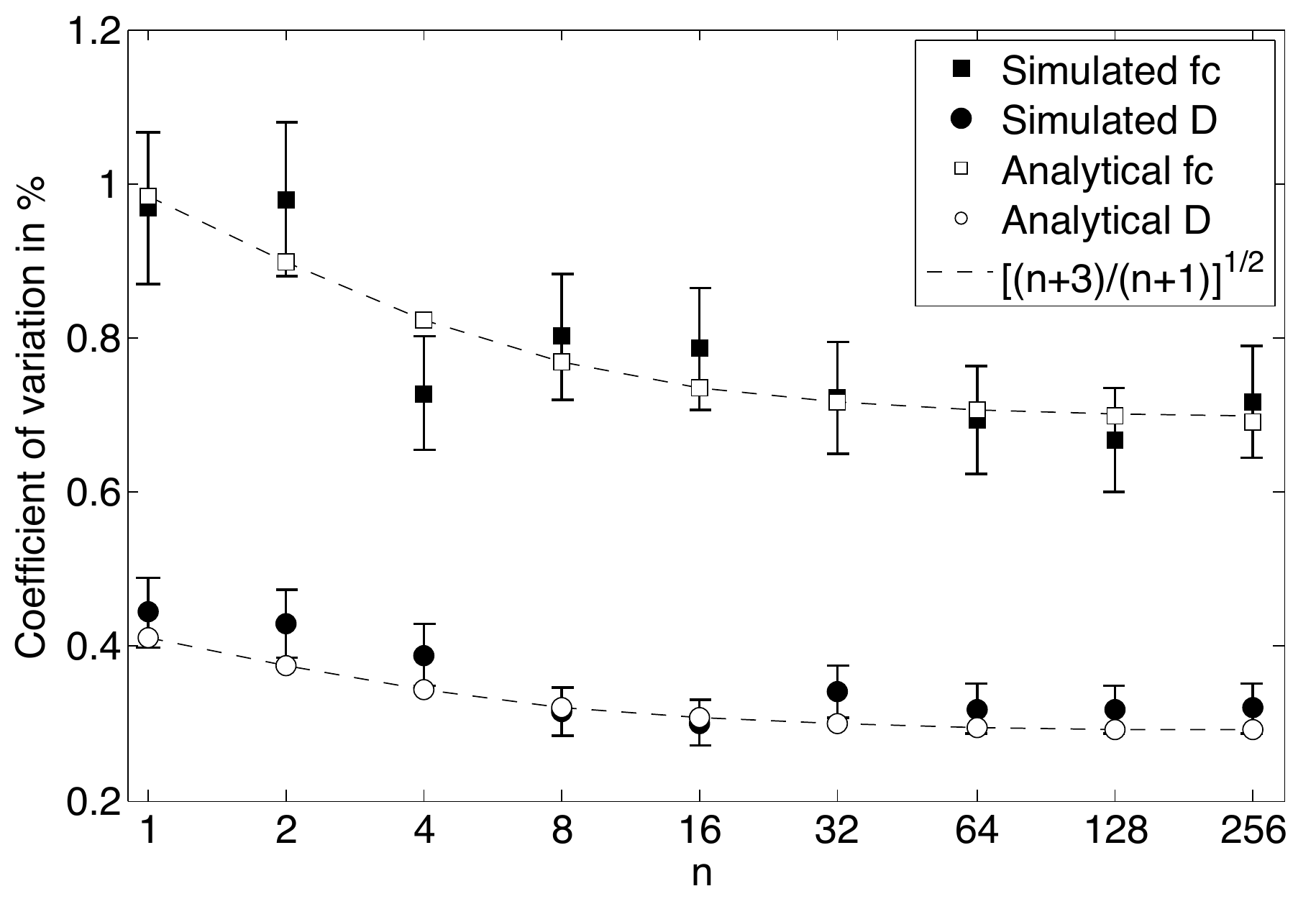}
\caption[]{\label{fig:errors}
Stochastic error on fit parameters as a function of number, $n$, of PSDs averaged over, for ML-fits of aliased Lorentzians using simplifying trick.
Compared to the dependence on the measurement time (Fig.~\ref{fig:N}), the error is only weakly dependent on $n$.
The sampling frequency was held fixed at   $\fsample=16,384$\,Hz, and the fit was done to the averaged  power spectra generated with  $N=262,144$ ($\Tmsr = 16$\,s), $\fc=500$\,Hz, and $D=0.46\,\mu$m/s$^2$.
Filled symbols show the coefficient of variation, $\sigma(X)/\la X\ra$, with $X=\fc, D$ determined using Eqs.~(\ref{eq:MLEfc})~and~(\ref{eq:MLED}), from 100 independent stochastic simulations.
Empty symbols show the theoretical expectation values from Eqs.~(\ref{eq:varfc})~and~(\ref{eq:varD}).
Dashed lines show the $\sqrt{(n+3)/(n+1)} \in [1:\sqrt{2}]$ scaling predicted from Eqs.~(\ref{eq:covabc})~and~(\ref{eq:covAB}).
}
\end{figure}

To find out how much of the available information we are putting to use, we can express the covariance matrix in terms of the Fisher information matrix $\mathcal{I} =N \tilde{\mathbf{S}}$ for the full (not the trick) MLE problem
\bea \label{eq:covabc}
	\mbox{cov}(a,b,c)  &\approx&  \mbox{cov}(\atrick,\btrick,\ctrick) = \frac{n+3}{n+1} \,   \mathcal{I} ^{-1}
\eea
from which we see that as $n$ increase, we approach the Cram\'er-Rao bound \cite{Rao} on the variance for an unbiased estimator, which is just $\mathcal{I}^{-1}$.
The main thing to notice, however, is that all the covariances in Eq.~(\ref{eq:covN}) are very small because $N$ is very large---the weak $n$ dependence (see Fig.~\ref{fig:errors}) is of no consequence in comparison.  
In an experiment, simply insert the fitted value for the power-spectrum as a best estimate for $P_f$ in $\tilde{\mathbf{S}}$.

\begin{figure}[ht]
 \includegraphics[width=8cm]{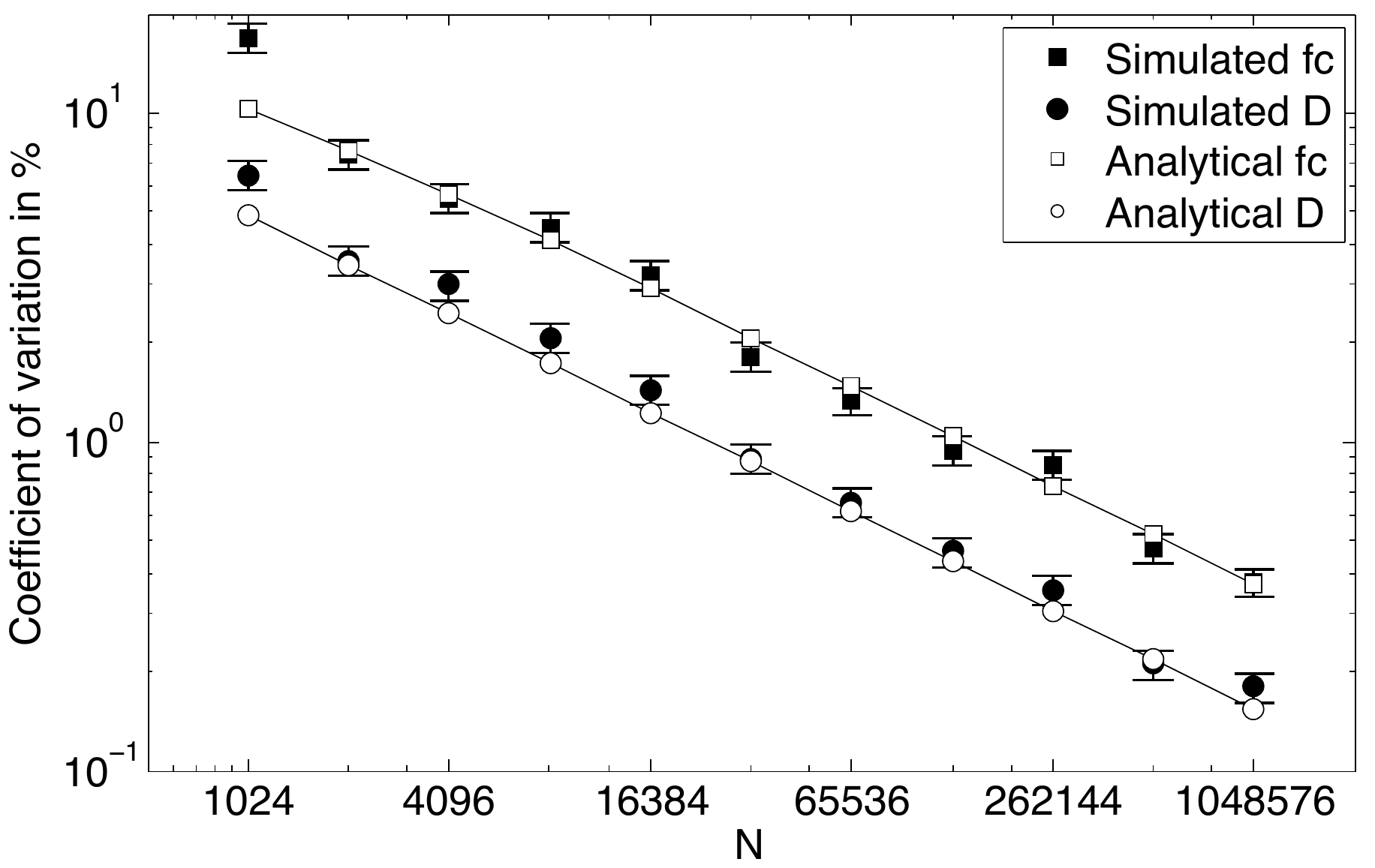}
\caption[]{\label{fig:N}
$1/\sqrt{N}$-dependence of stochastic error on fit parameters for ML-fits of aliased Lorentzians using simplifying trick.
An expected $1/\sqrt{N}$ behavior is easily made out on the double-logarithmic scale.
The sampling frequency was held fixed at $\fsample=16,384$\,Hz, and the fit was done to the average of $n=16$ power spectra generated with $\fc=500$\,Hz and $D=0.46\,\mu$m/s$^2$.
Filled symbols show the coefficient of variation, $\sigma(X)/\la X\ra$, with $X=\fc, D$ determined using Eqs.~(\ref{eq:MLEfc})~and~(\ref{eq:MLED}), from 100 independent stochastic simulations.
Empty symbols connected by lines show the theoretical expectation values from Eqs.~(\ref{eq:varfc})~and~(\ref{eq:varD}).
}
\end{figure}

\subsubsection{Results for Optical Tweezers}
For the non-aliased Lorentzian in Eq.~(\ref{eq:Lorentzian}) we then have, using the generic Eq.~(\ref{eq:differential})
\beq
	\sigma^2(\fc) =
	\frac{\fc^2}{4} \left[  \frac{ \la (\Delta a)^2\ra }{ a^2 } + \frac{  \la (\Delta b)^2\ra }{ b^2 }
	- 2 \frac{ \langle \Delta a \Delta b \rangle }{ ab }\right]
\eeq
and
\beq
	\sigma^2(D) = D^2 \frac{ \la (\Delta b)^2 \ra }{ b^2 }	\e,
\eeq
irrespective or the method used to estimate $a$, $b$, and $c$.
Actual numbers are found by inserting the estimates $(\atrick,\btrick,\ctrick)$.
The results for the aliased Lorentzian are given in Appendix~\ref{app:errorbarsaliasedOT}.

\subsubsection{Results for AFM cantilevers}
For the AFM, the variance of the three fit parameters are:
\bea
	\sigma^2(f_0) &=& \frac{f_0^2}{16} \left[ \frac{\la (\Delta a)^2 \ra}{a^2}
	 +  \frac{ \la (\Delta c)^2 \ra }{ c^2 }	- 2 \frac{  \la \Delta a \Delta c \ra }{ ac } \right]  \\
	 \nonumber\\
	\sigma^2(D) &=&
	\frac{D^4}{4 \pi^4} \left[ \la (\Delta a)^2 \ra f_0^{-4} + \la (\Delta b)^2 \ra
	+ \la (\Delta c)^2 \ra f_0^4 \right. \nonumber \\
	&+& 2 \left.  \la \Delta a \Delta b \ra f_0^{-2} + 2 \la \Delta a \Delta c \ra
	+ 2 \la \Delta b \Delta c \ra f_0^2 \right] \\
	\nonumber\\
	\sigma^2(G) &=&
	\frac{G^4}{c^2} \left[ \la (\Delta a)^2 \ra f_0^{-4} + \la (\Delta b)^2 \ra
	+ \la (\Delta c)^2 \ra (G^{-1} - f_0^2)^2 \right. \nonumber \\
	&+& 2 \left.  \la \Delta a \Delta b \ra f_0^{-2} + 2 \la \Delta a \Delta c \ra (1 - G^{-1} f_0^{-2} ) \right. \nonumber\\
	&+& 2\left. \la \Delta b \Delta c \ra (G^{-1} - f_0^2) \right]  \e,
\eea
where
\beq
	G = (2 \pi m / \gamma )^2 \e.
\eeq

\subsubsection{How good are the fits?}
The \emph{goodness of fit}, i.e., the {\em support\/} for the hypothesis that the fitted theory is correct, is \cite{Barford} the probability that a repetition of the experiment yields a data set with a smaller value for $p$.
A  calculation shows that for $K \gg 1$  the support, or backing, is
\beq \label{eq:support}
	B(s_0) = {\rm erfc}\left( |s_0-K| \sqrt{n/(2K)} \right)
\end{equation}
where ${\rm erfc}$ is the complementary error function, $s_0 \equiv \sum_f \Pbarexp_f/P_f$, and $K$ is the number of terms in the sum $s_0$.
We have assumed above that this number is much larger than the number of parameters fitted, hence equal to the number of degrees of freedom.
It is of order $10^4$--$10^6$ in our case, while 2--3 parameters are fitted.
When the sum $s_0 = K$, the expectation value for $s_0$, the backing is one but it rapidly drops to zero as $s_0$ becomes larger or smaller than $K$ .
The backing calculated for the fits shown in Fig.~\ref{fig:psd1} were, respectively,  0 (lsq fits; zero within the numerical precision of MatLab), 0.87 (ML-fit with simplifying trick), and 1.00 (rigorous numerical ML-fit; the first deviation from unity is in the 7th decimal place).
These numbers are stochastically varying because the PSD values are.
The backing for the rigorous numerical ML-fit will always be close to one because the stationarity conditions, see Eq.~(\ref{eq:costfunction}), for which the fit parameters are determined are virtually the same as $s_0 = K$.

\section{Summary and Conclusion}
\begin{figure}[ht]
 \includegraphics[width=8cm]{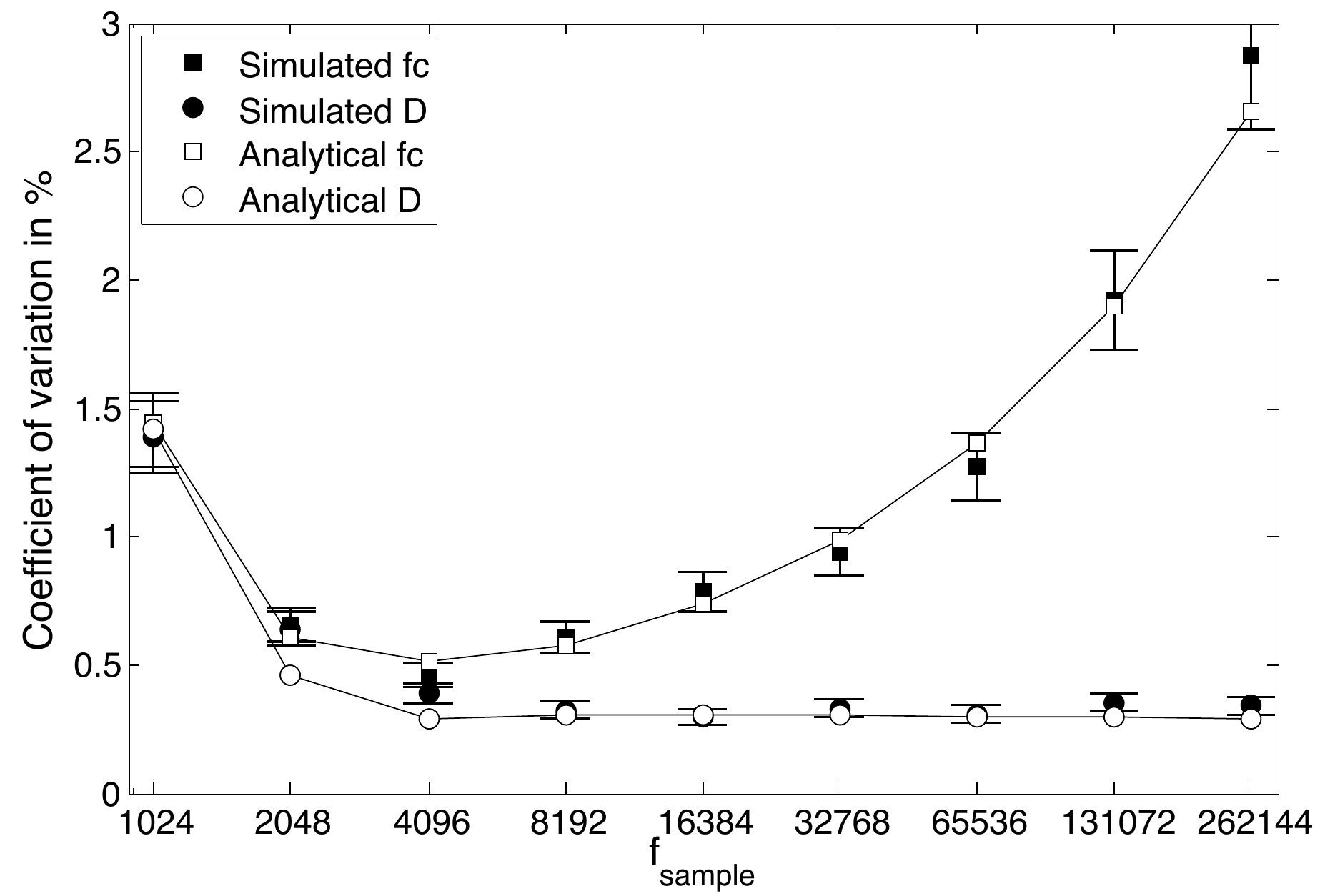}
\caption[]{\label{fig:errors2}
Stochastic error on fit parameters as a function of sampling frequency, $\fsample$, for ML-fits of aliased Lorentzians using simplifying trick.
A non-monotonic behavior is seen with minimum around $\fsample = 8 \fc$.
The number of acquired data points was held fixed at  $N=262,144$, and the fit was done to the average of $n=16$  power spectra generated with $\fc=500$\,Hz and $D=0.46\,\mu$m/s$^2$.
Filled symbols show the coefficient of variation, $\sigma(X)/\la X\ra$, with $X=\fc, D$ determined using Eqs.~(\ref{eq:MLEfc})~and~(\ref{eq:MLED}), from 100 independent stochastic simulations.
Empty symbols connected by lines show the theoretical expectation values from Eqs.~(\ref{eq:varfc})~and~(\ref{eq:varD}).}
\end{figure}

\begin{enumerate}
\item
A time series of $N$ coordinate values for an optically trapped microsphere or an AFM cantilever
doing Brownian motion, gives rise to $N$ power spectral values ($N/2+1$ distinct values) with signal-to-noise ratio 1 (Section~\ref{app:statprop}).
Parameters characterizing the power spectrum can consequently be determined with stochastic errors of order $1/\sqrt{N}$ in the aliased case (Section~\ref{sec:errorbars} and Appendix~\ref{app:errorbarsaliasedOT}).  For the non-aliased case see also Fig.~\ref{fig:aliasLSQ_variance}.
\item
For the purpose of displaying the experimental power spectrum,
its signal-to-noise value is reduced by a factor $\sqrt{n}$ by dividing the original data,
the time series of $N$ coordinates,
into $n$ equally long, non-overlapping subseries (Section~\ref{app:statprop}).
From these, $n$ experimental power spectra are calculated and averaged over.
This noise-reduced power spectrum covers the same frequency interval,
but the separation $\Delta f$ between consecutive points has increased by a factor $n$.
\item
Noise reduction trades resolution on the frequency axis for resolution on the power axis
in a manner that loses no information about the parameters characterizing the power spectrum. 
\item
The $1/\sqrt{n}$ scatter of experimental power spectral values should not be confused with
the expected stochastic error on parameter values characterizing the power spectrum;
the latter are of order $1/\sqrt{N}$ with only a weak $n$-dependence, as shown in Eq.~(\ref{eq:covN}), and Figs.~\ref{fig:N}~and~\ref{fig:errors}.
\item
Formulas given above, based on maximum likelihood estimates,  eliminate this issue and all further fitting
for the cases of the non-aliased  and aliased  Lorentzian power spectrum and for the non-aliased damped harmonic oscillator as model for an AFM cantilever (Section~\ref{app:simpletrick}).
Fitting has been done once-and-for-all and the results, including errorbars and a goodness-of-fit measure, are given by the formulas in Section~\ref{app:maxlikelihood} and Appendix~\ref{app:alias}.
\item
For other systems described by a linear Langevin equation driven by white or colored noise (Section~\ref{subsec:ubiq}),
one can determine parameters of the theoretical power spectrum
by fitting it to an experimental spectrum,
using weighted least-squares fitting
and experimental or theoretical error bars as weights,
which is computationally faster and more robust than MLE\@.
The resulting value for the diffusion coefficient $D$
should be corrected as described in Eqs.~(\ref{eq:Pbias}) and (\ref{eq:Pbiastheo}) respectively.
\item
Quite generally, beyond optical traps and AFM cantilevers, if the dependent variable (the data) and the squared weights are independent, then weighted least-squares yields unbiased results.
If not, the least-squares fit is typically biased, but this bias can often be removed altogether by a simple re-scaling of the fit-results (Section~\ref{sec:LSQ}).
\item 
To minimize the stochastic error on fit parameters we suggest setting $n=8$ or larger if using ML-fits with our simplifying trick or WLS with theoretical weights, see Eqs.~(\ref{eq:covabc})~and~(\ref{eq:covAB}) and Fig.~\ref{fig:errors}.  If WLS with experimental weights is used, we suggest $n=16$ or larger, see Fig.~\ref{fig:aliasLSQ_variance_n}. Also, for given $\fc$, $n$, and $N$, $\fsample = 8 \fc$ will minimize the stochastic error, see Fig.~\ref{fig:errors2}.  Typically however, $\Tmsr$ and $\fc$ are fixed and $\fsample$ should simply be set as high as meaningfully possible as $N$ is the most important factor in increasing the precision, see Fig.~\ref{fig:N}.  
\item
For optical tweezers, the highest precision is obtained by fitting over as much of the frequency range as can be captured by theory, including modification to the PSD due to hydrodynamics, optics, and instrumentation  \cite{Berg-Sorensen2004,Tolic-Norrelykke2006RSI,Schaeffer2007}.
If a certain level of precision is required it then becomes a matter of measuring long enough with the sampling frequency as high as experimentally possible (the minimum in Fig.~\ref{fig:errors2} is for \emph{fixed} $N$).
\item 
When doing least squares fitting we recommend the use of theoretical weights, Eq.~(\ref{eq:chi2theo}), if the standard deviation is known to be proportional to the expected value, because this minimizes the stochastic errors on the fit parameters, see Figs.~\ref{fig:aliasLSQ_variance}~and~\ref{fig:aliasLSQ_variance_n}. 
If any bias is present, that will also be smaller than the bias resulting from experimental weights, and can often be eliminated with the help of Eq.~(\ref{eq:ftrue}) or (\ref{eq:Pbiastheo}).
\end{enumerate}

\section{Acknowledgements}
We are thankful to Erik Sch\"affer for a critical reading of the manuscript.  SFN gratefully acknowledges financial support from the Carlsberg Foundation and the Lundbeck Foundation.  HF gratefully acknowledges financial support from the Human Frontier Science Program, GP0054/2009-C.
\clearpage

\appendix

\section{Monte Carlo simulation of the Einstein-Ornstein-Uhlenbeck theory of Brownian motion in a harmonic potential}
\label{app:EOU-MC}
In the case of non-negligible mass $m$, Eq.~(\ref{eq:NewtonMain}) is rewritten as two coupled first-order equations,
\beq
	\frac{\d}{\d t} \left( \begin{array}{c}  x(t) \\ v(t)  \end{array} \right) =
	-  \mathbf{M}   \left( \begin{array}{c}  x(t) \\ v(t)  \end{array} \right)
 	+ (2D)^{1/2}\frac{\gamma}{m}  \left( \begin{array}{c}  0 \\   \eta(t)  \end{array} \right)
	\enspace,
	\label{eq:OUtwo}
\eeq
where we have introduced the $2\times 2$ matrix
\beq
	\mathbf{M} = \left( \begin{array}{cc}  0 & -1 \\ \frac{\kappa}{m}  &  \frac{\gamma}{m} 	\end{array} \right)
	\enspace.
\eeq
Equation~(\ref{eq:OUtwo}) has the solution
\beq
 	\left( \begin{array}{c}  x(t) \\ v(t)  \end{array} \right) =
	(2D)^{1/2}\frac{\gamma}{m}  \int_{-\infty}^t  \d t' \, e^{- \mathbf{M}(t-t')}
 	\left( \begin{array}{c}  0 \\   \eta(t')  \end{array} \right)
	\enspace,
	\label{eq:OUtwosoln}
\eeq
from which follows that
\beq  \label{eq:OUiter}
 	\left( \begin{array}{c}  x_{j+1} \\ v_{j+1}  \end{array} \right) =
 	e^{- \mathbf{M}\Delta t}   \left( \begin{array}{c}  x_{j} \\ v_{j}  \end{array} \right)
	+  \left( \begin{array}{c}  \Delta x_j \\ \Delta v_j    \end{array} \right)
	\enspace.
\eeq
Here
\bea
	e^{- \mathbf{M}\Delta t} = \frac{1}{\lambda_+ - \lambda_-}
	\left( \begin{array}{cc} -\lambda_- c_+ + \lambda_+ c_- & -c_+ + c_- \\
	\lambda_+ \lambda_- (c_+ - c_-)  & \lambda_+ c_+ -\lambda_- c_-
	\end{array}\right)
\eea
where
\beq
	\lambda_{\pm} \equiv \frac{\gamma}{2m} \pm \sqrt{ \frac{\gamma^2}{4m^2} -
	\frac{\kappa}{m}  } \,\,\,,
	{\bf u_{\pm}} \equiv \left( \begin{array}{c} \mp1 \\ \pm\lambda_{\pm}   \end{array} \right)
\eeq
are the two eigenvalues and corresponding eigenvectors of $\mathbf{M}$,
\beq
	c_{\pm} = \exp(-\lambda_{\pm} \Delta t)
	\enspace,
\eeq
and we have introduced the notation
\beq
 	\left( \begin{array}{c}  \Delta x_j \\ \Delta v_j    \end{array} \right)
 	\equiv \Delta x_{+,j}  {\bf u_+ }
 	+   \Delta x_{-,j}  {\bf u_- }
\eeq
with
\beq
	\Delta x_{\pm,j}  \equiv (2D)^{1/2}
	\frac{\lambda_+ + \lambda_-}{\lambda_+ - \lambda_-}
	\int_{t_j}^{t_{j+1}}  dt' \, e^{-\lambda_{\pm} (t_{j+1}-t')}  \eta(t')
	\enspace.
\eeq
From Eq.~(\ref{eq:whitenoise}) follows that $\Delta x_{\pm,j}$ are two random lengths
drawn from Gaussian distributions with vanishing expectation value and known variances:
\bea
	\langle \Delta x_{\pm,j}\Delta x_{\pm,k} \rangle = &\sigma^2_{\pm}& \delta_{j,k}\enspace, \\
	&\sigma_{\pm}^2 &\equiv
	2D\left( \frac{ \lambda_+ + \lambda_- }{ \lambda_+ - \lambda_- } \right)^2
	\frac{1 - c_{\pm}^2}{2\lambda_{\pm}} \nonumber \enspace.
\eea
It also follows that $\Delta x_{+,j}$ and  $\Delta x_{-,j}$ are correlated with each other,
but uncorrelated with all $\Delta x_{\pm,k}$ for $j \neq k$:
\bea
 	\langle \Delta x_{+,j} \Delta x_{-,k} \rangle = &\sigma^2_{+-}& \delta_{j,k} \enspace, \\
	&\sigma_{+-}^2& \equiv
	2D\left( \frac{ \lambda_+ + \lambda_- }{ \lambda_+ - \lambda_- } \right)^2
	\frac{1 - c_+ c_-}{\lambda_+ + \lambda_-}  \nonumber \enspace.
\eea

From their known correlation follows, after some calculation, that they can be expressed in terms of two \emph{uncorrelated} Gaussian distributed
random numbers with unit variance, $\eta^{(a)}_j $ and  $\eta^{(b)}_j $, as
\bea
 	\lefteqn{ \left( \begin{array}{c}  \Delta x_j \\ \Delta v_j   \end{array} \right)  =}&&
 	\label{eq:discreteOUnoise} \\
 	&& \left(
     A_+ \left( \begin{array}{c}   -1  \\  \lambda_+   \end{array} \right)
 +   A_- \left( \begin{array}{c}    1  \\ -\lambda_-   \end{array} \right)
     \right) (1+\alpha)^{1/2} \, \eta^{(a)}_j
        \nonumber    \\
     &+& \left(
     A_+ \left( \begin{array}{c}   -1  \\  \lambda_+   \end{array} \right)
 -   A_- \left( \begin{array}{c}    1  \\ -\lambda_-   \end{array} \right)
     \right) (1-\alpha)^{1/2} \, \eta^{(b)}_j
    \enspace,  \nonumber
\eea
where we have introduced the notation
\beq
 	A_{\pm} = \frac{ \lambda_+ + \lambda_- }{ \lambda_+ - \lambda_- }
 	\sqrt{\frac{(1-c_{\pm}^2)D}{2\lambda_{\pm}}}
\eeq
and
\beq
 	\alpha = 2\frac{ \sqrt{\lambda_+ \lambda_-} }{ \lambda_+ + \lambda_- }
 	\frac{1-c_+ c_-}{\sqrt{(1-c_+^2)(1-c_-^2)}}
 	\enspace.
\eeq

So iteration of Eq.~(\ref{eq:OUiter}) with use of Eq.~(\ref{eq:discreteOUnoise}) generates a time series of positions $x_j$, which is sampled equidistantly in time with separation $\Delta t$ from the continuous-time solution to Eq.~(\ref{eq:NewtonMain}).
Since we use the exact analytical solution of Eq.~(\ref{eq:NewtonMain}) in the generation of this series, the finite value of $\Delta t$ causes no discretization error.
The only numerical errors associated with our solution are associated with the representation of real numbers on a computer, and, rather hypothetical, with the use of pseudo-random numbers.

\section{Aliased AFM Power Spectrum}
\label{app:aliasedAFM}
For the AFM, using the results from Appendix~\ref{app:EOU-MC}, we get for the aliased power spectrum:
\bea \label{eq:aliasedAFM}
	\langle \Pexp_k \rangle &=& \langle |\hat{x}_k^2| \rangle / \Tmsr  \nonumber \\
	&=&  \frac{ \sigma^2_+  \alpha_- + \sigma^2_-  \alpha_+ -2  \sigma^2_{+-}  \alpha_{+-} }
	{\alpha_+ \alpha_-} \Delta t
\eea
where
\bea
	\alpha_+ &=& 1+c_+^2 -2c_+ \cos(2\pi k/N) \\
	\alpha_- &=& 1+c_-^2 -2c_-\cos(2\pi k/N) \\
	\alpha_{+-} &=& 1+c_+ c_- -(c_+ + c_-) \cos(2\pi k/N)
\eea
and we have used that the discrete Fourier transform of $\eta_j^{(a)}$ and $\eta_j^{(b)}$ have the following characteristics
\bea
	\langle \hat{\eta}_k^{(a)*} \, \hat{\eta}_l^{(b)} \rangle &=& 0\\
	\langle \hat{\eta}_k^{(a)*} \, \hat{\eta}_l^{(a)} \rangle = \langle \hat{\eta}_k^{(b)*} \, \hat{\eta}_l^{(b)} \rangle &=& \Tmsr \Delta t \, \delta_{k,l} \enspace.
\eea
For completeness we note that the expression in Eq.~(\ref{eq:aliasedAFM}) has as limiting expression Eq.~(\ref{eq:Palias2}) when the mass vanishes:
\beq
	\lim_{m \rightarrow 0} \langle \Pexp_k \rangle = \frac{\sigma^2_-}{\alpha_-} \Delta t
\eeq
as is seen by inspection.

\section{Maximum Likelihood Estimation for aliased power spectra}
\label{app:alias}
In an OT experiment, the time-series of bead positions $x(t)$ is obtained by sampling the continuous output from the photodiode at discrete times $t_j = j\Delta t$, $\Delta t = 1/\Tmsr$.
Applying the discrete Fourier transform to $x(t)$ we find \cite{Berg-Sorensen2004} that the expectation value for the aliased power spectrum can be written in the form:
\beq \label{eq:Palias2}
	P_k^{\rm alias} = \frac{1}{A + B \cos(2\pi k/N)} \enspace,
\eeq
where $A$ and $B$ are related to $\fc$ and $D$ through
\bea
	\fc &=& \frac{ \fsample} {2\pi} \, u  \label{eq:MLEfc} \\
	D &=& \frac{ \fsample^2 }{ A \tanh(u) }  \, u \label{eq:MLED} \enspace , \\
	u &=& \cosh^{-1}(-A/B) \label{eq:u}\e.
\eea

By inserting Eq.~(\ref{eq:Palias2}) in Eq.~(\ref{eq:costfunction}) the stationarity conditions ($\partial_A \mathcal{F} = \partial_B \mathcal{F} = 0$) are seen to be
\bea	
	\sum_f  \Pbarexp_f   &=& \sum_f P_f  \label{eq:nonlinalias1}\\
	\sum_f   \cos(2\pi k/N) \Pbarexp_f   &=& \sum_f  \cos(2\pi k/N) P_f \label{eq:nonlinalias2}\enspace.
\eea
We now repeat the trick introduced in Section~\ref{app:simpletrick} to turn Eqs.~(\ref{eq:nonlinalias1})~and~(\ref{eq:nonlinalias2}) into expressions linear in $A$ and $B$
\beq \label{eq:Rur}
	\underbrace{
	\left( \begin{array}{cc}
	R_{0,2} & R_{1,2}   \\
	R_{1,2} & R_{2,2}
	\end{array}\right)
	}_{\mathbf{R}}
	\underbrace{
	\left(\begin{array}{c}
	A \\
	B
	\end{array}\right)
	}_{\vec{u}}
	\approx (1+1/n)
	\underbrace{
	\left(\begin{array}{c}
	R_{0,1}  \\
	R_{1,1}
	\end{array}\right)
	}_{\vec{r}} \e,
\eeq

that are solved to give
\bea
	A &\approx& \Atrick \equiv  \frac{n+1}{n}\frac{R_{0,1}R_{2,2}-R_{1,1}R_{1,2}} {R_{0,2}R_{2,2}-R_{1,2}^2}
	\label{eq:MLEA} \\ && \nonumber\\
	B &\approx& \Btrick \equiv \frac{n+1}{n}\frac{R_{0,2}R_{1,1}-R_{0,1}R_{1,2}} {R_{0,2}R_{2,2}-R_{1,2}^2} \label{eq:MLEB} \enspace.
\eea
where we have introduced the (aliased) statistics
\beq \label{eq:Rpq}
	R_{p,q} = \frac{1}{K}\sum_k \cos^p(2\pi k/N)\,  \Pbarexpq_k  \enspace.
\eeq

We do not attempt here to give the aliased results for $f_0, D$, and $G$ from the AFM case:
To avoid the aliasing of high frequency noise to the lower frequencies of interest,  a high sampling frequency is often used when acquiring AFM data.
However, only the region around $f_0$ is well captured by the dampened harmonic oscillator theory and therefore no more than this region is fitted.
Since the aliased expressions only deviate substantially from the non-aliased ones at high frequencies, and because the non-aliased expressions are much simpler, we only treated the non-aliased MLE for the AFM here.

\section{Covariance Matrix}
\label{app:covariance}
This appendix derives the results that are used in Section~\ref{sec:errorbars} and Appendix~\ref{app:errorbarsaliasedOT}.
To calculate the covariance matrix we look at the response of the estimated fit parameters to fluctuations in the statistics $S_{p,q}$. The calculations go through unchanged for the aliased Lorentzian (see above) with statistics $R_{p,q}$.
First, we note that we can write each term in Eq.~(\ref{eq:Svs3})
\beq \label{eq:Svs}
	\mathbf{S} \vec{v} \approx \mathbf{S} \hat{\vec{v}} \equiv 
	 \mathbf{S} \left( \begin{array}{c} \atrick \\ \btrick \\ \ctrick \end{array}\right) 
	 \equiv	( 1 + 1/n ) \vec{s}
\eeq
as the sum of its ``true underlying'' value and a fluctuation ($\Delta \! \mathbf{S}$ and $\Delta\vec{s}$) or a response ($\Delta \vec{v}$) to fluctuations
\bea
	\mathbf{S} &=&  \la \mathbf{S} \ra + \Delta \! \mathbf{S}\\
 	\hat{\vec{v}} &=& \vec{v} + \Delta \vec{v} \label{eq:vecv}\\
	\vec{s} &=&  \la \vec{s} \ra + \Delta \vec{s}
\eea
where the elements in $\Delta \! \mathbf{S}$ are
\beq
	 \Delta S_{p,q} =\frac{1}{K}  \sum_f f^{2p} \left[ \Pbarexpq - \la \Pbarexpq \ra \right]
\eeq
To first order in the fluctuations we thus have
\beq \label{eq:deltav}
	\Delta \vec{v} =  \left( \begin{array}{c} \Delta a\\ \Delta b\\ \Delta c \end{array}\right) =
	\la\mathbf{S}\ra^{-1} \left( \frac{n+1}{n} \, \Delta \vec{s} - \Delta \! \mathbf{S} \vec{v} \right)
\eeq
where we notice that to first order $\la \Delta \vec{v} \ra =0$ which, however, does not mean that this is an unbiased estimator as shown below.
The covariance matrix $\la \Delta\vec{v} \otimes \Delta \vec{v} \ra$ as given in Eq.~(\ref{eq:covariance}) then follows after some calculation using Eqs.~(\ref{eq:Pabc}) and (\ref{eq:qmoment}).

We emphasize here, that the above calculations were to first order in $1/\sqrt{K}$, with $K$ the number of terms in the statistics $S_{p,q}$:
Whereas the relative size of an individual fluctuation in the power spectral value $\Delta \Pexp$ is independent of $K$, the relative sizes of the overall fluctuations in the sums $\Delta S_{p,q}$ go to zero as $1/\sqrt{K}$. Since $K$ is typically of the order $10^4$--$10^6$, this first-order approximation is very good.

\section{Error-bars for the aliased Lorentzian}
\label{app:errorbarsaliasedOT}

The error-bars on $\fc$ and $D$ are calculated as before, using Eq.~(\ref{eq:differential}) for the variance:
\bea \label{eq:varfc}
	\sigma^2(\fc) &=& \frac{ \fsample^2 A^2 }{ 4 \pi^2 (A^2 - B^2) } \left[
	\frac{ \la (\Delta A)^2\ra }{ A^2 } + \frac{ \la (\Delta B)^2\ra }{ B^2} \right. \nonumber \\
	&-& \left. 2 \frac{ \la \,\Delta A \Delta B \ra }{ AB } \right]
\eea
and
\bea \label{eq:varD}
	\sigma^2(D) &=& (\partial_A D)^2 \la (\Delta A)^2\ra + (\partial_B D)^2 \la (\Delta B)^2\ra \nonumber \\
	&+& 2 \partial_A D \, \partial_B D \la \,\Delta A \Delta B \ra
\eea
where
\bea
	\partial_A D &=& \frac{D}{A} \left( \frac{B^2}{A^2-B^2} - 1 - \frac{A}{u \sqrt{A^2 -B^2} }\right) \\
	\nonumber\\
	\partial_B D &=& \frac{D}{B} \left( \frac{A}{u \sqrt{A^2 -B^2} } - \frac{B^2}{A^2-B^2}\right)\e,\\
	\nonumber
\eea
and $u$ is given in Eq.~(\ref{eq:u}).
The covariance matrix is calculated as before, giving
\bea \label{eq:covAB}
	\mbox{cov}(A,B) &\equiv& \left( \begin{array}{cc}
	\la (\Delta A)^2 \ra & \la \Delta A \Delta B \ra \\
	\la \Delta A \Delta B \ra & \la (\Delta B)^2 \ra
	\end{array}\right) 	\nonumber \\
	&\approx&  \frac{1}{N}\frac{n+3}{n+1}  \tilde{\mathbf{R}} ^{-1} \e,
\eea

where $\tilde{\mathbf{R}}$ is a matrix with the same structure as $\mathbf{R}$, see Eq.~(\ref{eq:Rur}), but with the experimental values $\Pbarexp$ replaced by the theoretical (in practice the fitted) values $P_f$ in all the statistics Eq.~(\ref{eq:Rpq}).

We tested the above analytical results by comparing to the results of simulations:
Multiple independent position time-series for a mass-less particle diffusing in a harmonic potential were created using the methods given in \cite{Berg-Sorensen2004}.
The resulting power spectra were fitted using Eqs.~(\ref{eq:MLEfc}--\ref{eq:Rpq}).
Since we simulate a stochastic process there is scatter in the fitted parameters and it is this scatter that we compare to the results given in Eqs.~(\ref{eq:varfc})~and~(\ref{eq:varD}).
The results are shown in Figs.~\ref{fig:N} and \ref{fig:errors}.
Figure~\ref{fig:N} shows the expected $1/\sqrt{N}$ scaling, whereas Fig.~\ref{fig:errors} shows a $\sqrt{n+3}/\sqrt{n+1}$ scaling.

Figure~\ref{fig:errors2} shows that the optimal tradeoff between precision and amount of data acquired seems to manifest itself at a sampling frequency roughly eight times the corner frequency; any slower than this leads to large errors in both parameters because the PSD essentially reduces to the ratio of $D$ to $\fc$, i.e., two parameters are used to fit a single constant.  Sampling much faster than $\fc$, but keeping the number of acquired data points fixed, has no effect on the error on $D$ but is detrimental for $\fc$ since there is progressively less information about $\fc$ at larger frequencies.
The effect on the precision of increasing $n$ is small but positive; compared to the effect of $N$ and $\fsample$  it can be ignored (after including it as described in Eqs.~(\ref{eq:MLEA}) and (\ref{eq:MLEB})).

For comparison, the error-bars on the fit parameters, as a function of cut-off frequency $f_{\rm max}$, from a least-squares fit are shown in Fig.~\ref{fig:aliasLSQ_variance}:  The average of $n=16$ synthetic PSDs were fitted by minimizing Eq.~(\ref{eq:chi2sigma}) with the data-points weighted by the standard deviation of the $n$  PSDs.
Also shown is a LSQ fit where the weights are kept constant; this is the kind of fitting performed by primitive LSQ routines.
For a detailed discussion of how the stochastic error depends on the fitting range $[f_{\rm min} : f_{\rm max} ]$ the reader is referred to section VIII in \cite{Berg-Sorensen2004}.

\section{Bias of the MLEs}
\label{app:bias}
Obviously, we must be paying a price somewhere
for turning a non-linear problem into a linear one with our little trick,
or else we would have turned a non-linear problem into an exactly solvable mathematical problem.
That is sometimes done, but not here:
The trick works through an approximation,
and the resulting approximate estimator is biased,
which means that its expectation value is different from the true value
of the quantity it estimates.
So on the average it misses the correct result.
Bias is systematic error on averages.
That is the \emph{nature} of the price we pay.
Fortunately, it is negligible in size, as we demonstrate now.

For the non-aliased Lorentzian and AFM we find, by expanding Eq.~(\ref{eq:Svs}) to first order in $\Delta \vec{v}$ and \emph{second} order in $\Delta S_{p,q}$
\bea  \label{eq:bias}
	\lefteqn{\la \Delta \vec{v} \ra = \left( \begin{array}{c} \la \Delta a \ra \\ \la\Delta b \ra \\ \la \Delta c \ra \end{array} \right) } &&\\
	&=& \la \mathbf{S} \ra ^{-1} \left[
	 \la \Delta \mathbf{S} \la\mathbf{S}\ra^{-1} \Delta \mathbf{S} \ra \vec{v}
	- \frac{n+1}{n} \la \Delta  \mathbf{S} \la\mathbf{S}\ra^{-1} \Delta \vec{s} \ra 	\right] \nonumber \e.
\eea
For the non-aliased Lorentzian this expression can be reduced to
\bea
	\left( \begin{array}{c} \la \Delta a \ra \\ \la\Delta b \ra \end{array} \right)
	=\frac{2}{N} \, \frac{n+2}{n+1} 	\,
	\left(\tilde{S}_{0,2}\tilde{S}_{2,2} - \tilde{S}_{1,2}^2\right)^{-2} &&\\
	\left(\begin{array}{cc}
	\tilde{S}_{2,2} & -\tilde{S}_{1,2}\\
 	-\tilde{S}_{1,2} & \tilde{S}_{0,2}
	\end{array} \right)
	\left(\begin{array}{ccc}
	\tilde{S}_{2,3} & -2\tilde{S}_{1,3} & \tilde{S}_{0,3} \\
 	\tilde{S}_{3,3} & -2\tilde{S}_{2,3} & \tilde{S}_{1,3}
	\end{array} \right)
	\left( \begin{array}{c}
	\tilde{S}_{0,2} \\ \tilde{S}_{1,2} \\ \tilde{S}_{2,2}
	\end{array} \right)  &&
	\nonumber
	\eea
where
\beq \label{eq:Stilde}
	\tilde{S}_{p,q} = \frac{1}{K} \sum_f f^{2p} \la \Pbarexp \ra^{q} = \frac{1}{K} \sum_f f^{2p} P_f^q  \e.
\eeq
That is, the bias is proportional to $1/N$ which is a very small number, and displays a weak $n$-dependence.
From these expressions we find the bias on $\fc$ and $D$ to be
\bea
	\la \Delta \fc \ra &=& \frac{\fc}{2} \, \left( \frac{ \la \Delta a \ra }{ a } - \frac{ \la \Delta b \ra }{ b } \right) \\
	\la \Delta D \ra &=& -D  \, \frac{ \la \Delta b \ra }{ b } \e.
\eea
We only know the true values of $\fc$ and $D$ in simulations, in an experiment the best estimate of the true value would be the fitted value.

\begin{figure}[ht]
 \includegraphics[width=8cm]{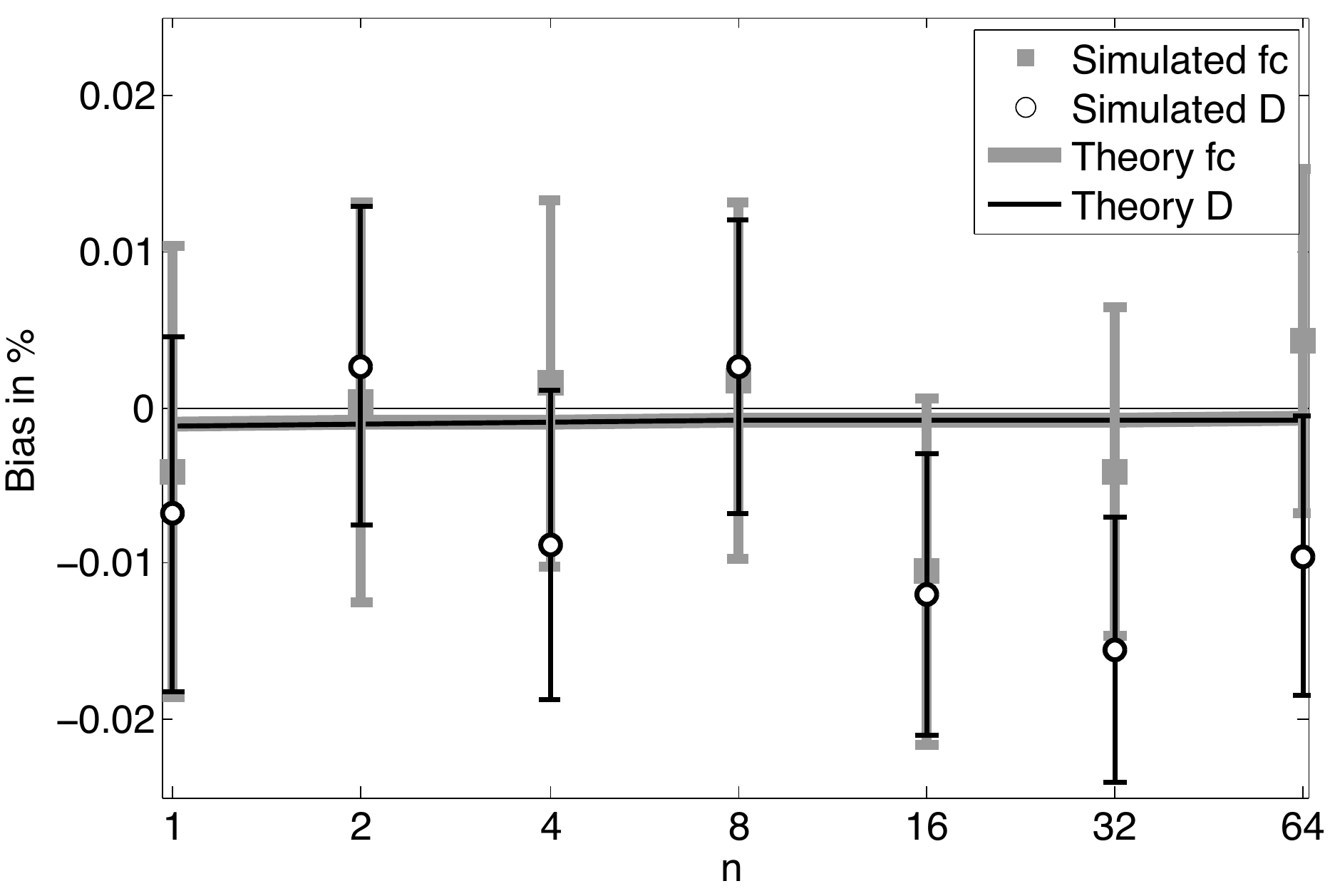}
\caption[]{\label{fig:bias}
Bias of the parameter estimates for $\fc$ (grey squares) and $D$ (white circles), from ML-fits with our trick, of the aliased Lorentzian.  Error-bars are standard errors on the mean.
The point of this figure is only to show that the bias is indeed very small and can be completely ignored.
Theoretical expectation value for the bias as given in Eq.~(\ref{eq:biasfc}) (grey line) and Eq.~(\ref{eq:biasD}) (black line) are less than $1/1,000$ of a percent for these settings.
Bias was measured as the difference between the average of 1,000 determinations of the fit parameters and the known input values $\fc=500$\,Hz and $D=0.46\,\mu$m/s$^2$.
Simulations were run with  $N=2^{20}$ and $\fsample=2^{12}$\,Hz---values chosen to minimize the stochastic errors.  Data were treated with $n$ non-overlapping Hann windows before calculation of the PSD.}
\end{figure}

For the aliased Lorentzian we find in an analogous manner
\bea
	\la \Delta \fc \ra &=& \frac{ \fsample }{ 2 \pi } \frac{ A }{ \sqrt{ {A}^2 - {B}^2} }
	\left[ \frac{ \la \Delta B \ra}{ B } - \frac{ \la \Delta A \ra}{ A } \right] \label{eq:biasfc} \\
	\nonumber \\
	\la \Delta D \ra &=& -D \left[ \left( \frac{ {B}^2 }{ {A}^2 - {B}^2  } +
	\frac{ A }{ \sqrt{ {A}^2 - {B}^2}  }  \right) \right. \nonumber\\
	&&\left. \left( \frac{ \la \Delta A \ra }{ A } + \frac{ \la \Delta B \ra }{ B } \right)
	- \frac{ \la \Delta A \ra }{ A } \right]  \label{eq:biasD} \e,
\eea
where $\la \Delta A \ra$ and $\la \Delta B \ra$ are found from Eq.~(\ref{eq:bias}) by everywhere replacing
 $\tilde{S}$ with $\tilde{R}$.
The result of a numerical test of the above relations is shown in Fig.~\ref{fig:bias}.

\begin{thebibliography}{10}

\bibitem{Neuman2004}
K.~C. Neuman and S.~M. Block, Review of Scientific Instruments {\bf 75},  2787
  (2004).

\bibitem{Neuman2007}
K.~C. Neuman, T. Lionnet, and J.-F. Allemand, Annu Rev Mater Res {\bf 37},  33
  (2007).

\bibitem{Neuman2008}
K.~C. Neuman and A. Nagy, Nat Meth {\bf 5},  491  (2008).

\bibitem{Moffitt2008}
J.~R. Moffitt, Y.~R. Chemla, S.~B. Smith, and C. Bustamante, Annu Rev Biochem
  {\bf 77},  205  (2008).

\bibitem{Perkins2009}
T.~T. Perkins, Laser {\&} Photon. Rev. {\bf 3},  203  (2009).

\bibitem{Hutter1993}
J.~L. Hutter and J. Bechhoefer, Rev Sci Instrum {\bf 64},  1868  (1993).

\bibitem{Walters1996}
D. Walters {\it et~al.}, Rev Sci Instrum {\bf 67},  3583  (1996).

\bibitem{Sader1998}
J. Sader, Journal of applied physics {\bf 84},  64  (1998).

\bibitem{Berg-Sorensen2004}
K. Berg-S{\o}rensen and H. Flyvbjerg, Review of Scientific Instruments {\bf
  75},  594  (2004).

\bibitem{Tolic-Norrelykke2006RSI}
S.~F. Toli\'c-N{\o}rrelykke {\it et~al.}, Review of Scientific Instruments {\bf
  77},  103101  (2006).

\bibitem{Berg-Sorensen2003}
K. {Berg-S{\o}rensen}, L. {Oddershede}, E.-L. {Florin}, and H. {Flyvbjerg},
  Journal of Applied Physics {\bf 93},  3167  (2003).

\bibitem{Berg-Sorensen2006}
K. Berg-S{\o}rensen {\it et~al.}, Review of Scientific Instruments {\bf 77},
  063106  (2006).

\bibitem{Schaeffer2007}
E. Sch\"affer, S.~F. N{\o}rrelykke, and J. Howard, Langmuir {\bf 23},  3654
  (2007).

\bibitem{Aitken1935}
A.~C. Aitken, Proceedings of the Royal Society of Edinburgh {\bf 55},  42
  (1935).

\bibitem{Rao}
C.~R. Rao, {\em {Linear Statistical Inference and Its Applications}} (Wiley,
  New York, New York, 1973).

\bibitem{Barford}
N.~C. Barford, {\em {Experimental Measurements: Precision, Error and Truth}},
  2nd  ed. (John Wiley \& Sons, 1990).

\bibitem{footnote}
Here and below, we used the following results that are easy to verify by direct calculation or by consulting a statistics text book:  If two independent stochastic variables are normally distributed $X,Y \sim \mathcal{N}(0,\sigma^2/2)$, then their individual squares are gamma distributed $X^2,Y^2 \sim \Gamma( \frac{1}{2},\frac{1}{\sigma^2} )$, and the sum of their squares is exponentially distributed $Z =  X^2 + Y^2  \sim  \Gamma(1,\frac{1}{\sigma^2}) = E(\frac{1}{\sigma^2})$.  Finally, averaging over $n$ independent, exponentially distributed variables returns a gamma distributed variable $\frac{1}{n} \sum_{i=1}^n Z_i \sim \Gamma(n,\frac{n}{\sigma^2})$.

\end{thebibliography}

\end{document}